\documentclass[11pt]{article}

\usepackage{amsmath}  
\usepackage{subdepth}
\usepackage[usenames,dvipsnames]{color}

\usepackage{amsfonts} 
\usepackage{eurosym}

\usepackage{natbib}
\usepackage{a4}
\usepackage{epsf}
\usepackage{graphicx}
\usepackage{rotating}
\usepackage{tabularx}

\usepackage[latin1]{inputenc}

\usepackage{bm}
\renewcommand{\vec}[1]{\bm{#1}}

\newcommand{\beq}{\begin{equation}}
\newcommand{\eeq}{\end{equation}}

\numberwithin{equation}{section} 

\newcommand{\wcaption}[2]{\caption{#1}\label{#2}}

\newcommand{\rit}{{\rm I\!R}}

\newcommand{\ttf}[1]{{\ttfamily{#1}}}

\newcommand{\G}[2]{{\cal G }\left(#1,#2\right)}
\newcommand{\D}{{\cal D }}

\begin{document}


\title{Spline approximations to \\ conditional Archimedean copula}
\author{Philippe Lambert
\footnote{Institut des sciences humaines et sociales, M\'ethodes
quantitatives en sciences sociales, Universit\'e de Li\`ege, Li\`ege,
Belgium. Email: p.lambert@ulg.ac.be.}
\footnote{Institut de statistique, biostatistique et sciences actuarielles (ISBA), Universit\'e catholique de Louvain,Louvain-la-Neuve, Belgium.}
}


\date{November 13, 2013} 

\maketitle

\begin{abstract}
  We propose a flexible copula model to describe changes with a
  covariate in the dependence structure of (conditionally
  exchangeable) random variables. The starting point is a spline
  approximation to the generator of an Archimedean copula.
  Changes in the dependence structure with a covariate $x$ are
  modelled by flexible regression of the spline coefficients on $x$.
  The performances and properties of the spline estimate of the
  reference generator and the abilities of these conditional models to
  approximate conditional copulas are studied through simulations.
  Inference is made using Bayesian arguments with posterior
  distributions explored using importance sampling or adaptive MCMC
  algorithms. The modelling strategy is illustrated with two
  examples. 

\end{abstract}



\section{Introduction}
\citet{SklarA:fonc:59}
has proved that any distribution $H(y_{1},\ldots,y_{p})$ with marginal distributions
$F_{j}(y_{j})~(j=1,\ldots,p)$ can be written as 
\begin{eqnarray}
H(y_{1},\ldots,y_{p})=C(F_{1}(y_{1}),\ldots,F_{p}(y_{p})), \label{lambert:CopulaEq1}
\end{eqnarray}
where $C$ denotes a distribution function (named {\em copula}) on
$(0,1)^{p}$ with uniform margins.  If the margins are continuous, then
$C$ is unique.  Conversely, if $C$ is a copula and $F_{j}(\cdot)$ are
distribution functions, then (\ref{lambert:CopulaEq1})
defines a multivariate distribution with marginal distributions
$F_{j}$ ($j=1,\ldots,p$).

In most practical applications where copula are used, the marginal
distributions and their potential link with covariates $\bold{x}$ are
investigated in a first step, yielding marginal fitted quantiles,
$\hat{u}_{j|x}=\hat{F}_{j}(y_{j}|\bold{x})~(j=1,\ldots,p)$. A
parametric copula is then selected to describe the dependence
structure of the fitted quantiles. That copula is usually assumed to
be independent of the covariates as if the strength of association
between the margins did not change with the unit characteristics. Then,
(\ref{lambert:CopulaEq1}) becomes
\begin{eqnarray}
H(y_{1},\ldots,y_{p}|\bold{x})=C(F_{1}(y_{1}|\bold{x}),\ldots,F_{p}(y_{p}|\bold{x})), \label{lambert:CopulaEq2}
\end{eqnarray}

\begin{figure}[bt!]\centering
 \includegraphics[width=12cm]{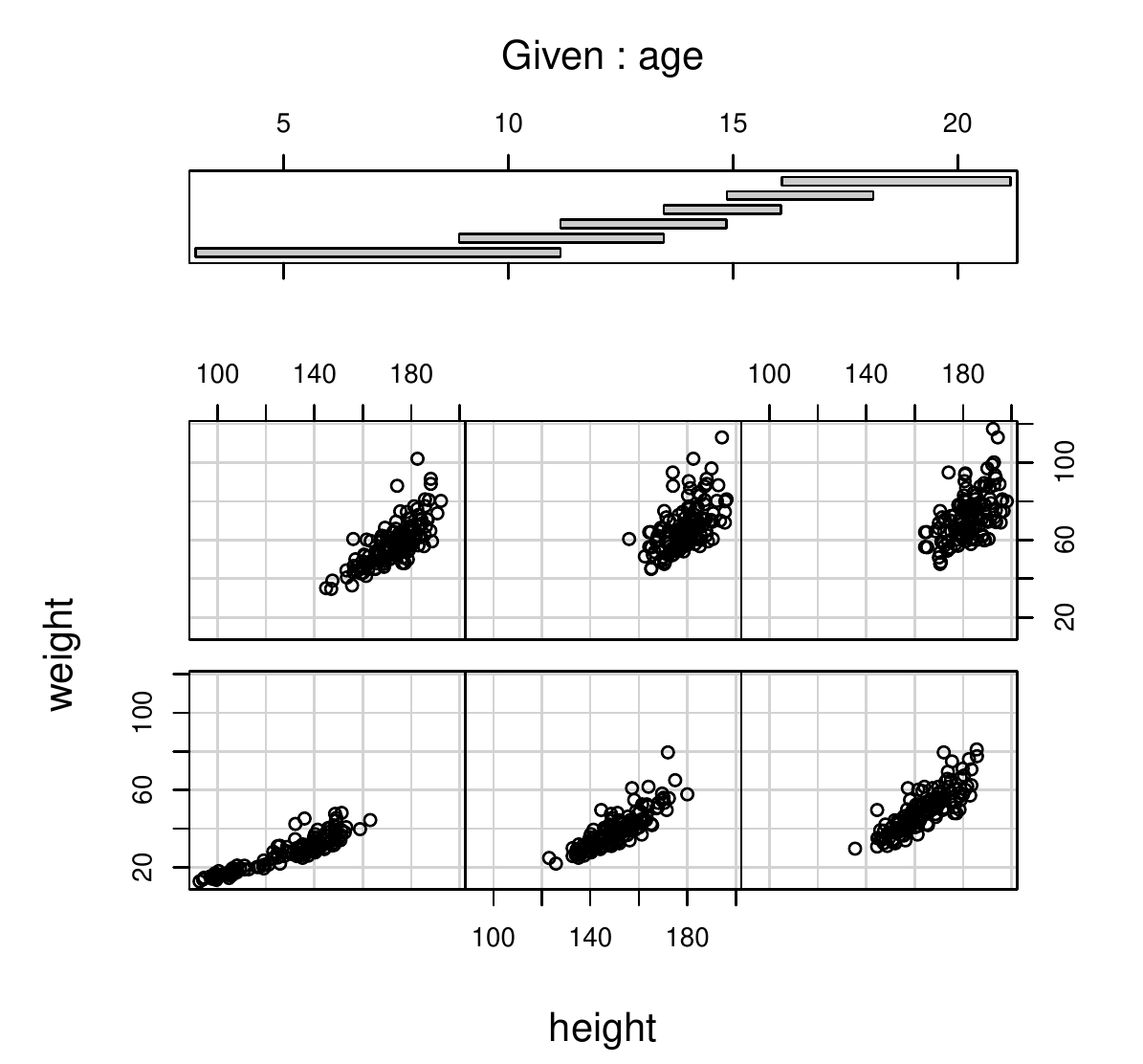}
 \caption{\label{lambert:DutchBoysCoplot}
   Growth dataset (for 441 Dutch boys): conditional scatterplot of
   weight versus height for different age classes.}
\end{figure}

Unfortunately, the previous modelling assumption is not always
realistic, as shown on Fig.~\ref{lambert:DutchBoysCoplot} where the
scatterplot of weight (in kg) and height (in cm) of young boys is
given for different age classes.
Indeed, one probably notices that the link between the marginal
quantiles is getting looser as age increases.  When the selected
copula $C_{\theta}$ is parametric, one can let the copula parameter
(and, hence, the strength of association) change with covariates,
yielding
\begin{eqnarray}
H(y_{1},\ldots,y_{p}|\bold{x})=C_{\theta(\bold{x})}(F_{1}(y_{1}|\bold{x}),\ldots,F_{p}(y_{p}|\bold{x})), \label{lambert:CopulaEq3}
\end{eqnarray}
An early example of that can be found in \citet{LambertP:acop:02}
where the effect of an antidepressant on blood pressures and heart
rate were studied in a longitudinal setting. Besides the effects of
covariates on the marginal distributions of these three responses,
their strengths of association were also allowed to change with sex
and the presence of drug in the plasma. The same idea was used in a
financial context by \citet{Patton:2006uy}
where the name {\em conditional copula} for $C_{\theta(\bold{x})}$ was
coined.

Nonparametric versions are desirable to suggest or to validate
parametric specifications, or even as a substitute for these
models. This is a growing topic of research in the copula literature.
\citet{Hafner:2010jr} and \citet{Acar:2011tc} use local likelihood to
estimate changes with a covariate in the dependence parameter of a
parametric copula, while \citet{CraiuSabeti:2012} explore the use of
cubic splines.
Nonparametric estimates for the conditional copula
were also proposed and studied by \citet{Gijbels:2011},
\citet{Veraverbeke:2011gz} and \citet{Abegaz:2012he} where a local
kernel-weighted pseudo likelihood is used to estimate
$\theta(\bold{x})$ in a local polynomial approach. Extensions to
multivariate or even functional covariates can be found in
\citet{Gijbels:2012ke}.

Here, we explore the possibility to use B-splines to specify a copula
'non-parame\-trically' and to model its evolution with a covariate.
The plan of the paper is as follows. In Section
\ref{CopulaSpline:Sect}, we extend the work of \citet{Lambert:2007vo}
and \citet{Vandenhende:2005ug} to provide a smooth estimate for the
generator of an Archimedean copula. Its properties are revealed by a
simulation study in Section \ref{CopulaSpline:SimulationStudy:Sect}
with excellent results already available with small sample sizes.
Flexible power transforms of the copula generator are introduced in
Section \ref{FlexPowerFamily:Sect} to model smooth changes of the
copula with covariates. The abilities of the {\em flex-power} and of
the {\em additive conditional spline Archimedean copula families} to
model such changes are studied in Sections
\ref{AdditiveCondSplineFamily:Sect} and \ref{FlexPowerFamily:Sect}. We
conclude the paper with applications in Section \ref{Application:Sec}
followed by a discussion.

\section{Spline estimation of an Archimedean
  copula} \label{CopulaSpline:Sect}

We shall restrict our attention to the estimation of Archimedean
copulas.  A (bivariate) copula is said {\em Archimedean}
\citep{GenestC:thej:86} 
if it can be written as
$$C(u,v) = \varphi^{-1}\left({\varphi(u)+\varphi(v)}\right)$$
where $\varphi(\cdot)$ is a decreasing and convex function (named the
{\em generator}) taking values on $(0,1)$ and such that
$\varphi(0^{+})=+\infty$ and $\varphi(1)=0$. It is symmetric in $u$
and $v$, and characterized by that univariate function.  The strength
of association between the two variates can be quantified using
e.g.~Spearman's rho or Kendall's tau. They can be computed from the
generator with the latter given by
\begin{eqnarray}
\tau = 1 + 4\int_0^1 {\lambda(s)} ds, \label{KendallTauAndLambda:Eqn}
\end{eqnarray}
where $\lambda(\cdot) = \varphi(\cdot) / \varphi'(\cdot)$. 

Various parametric proposals $\varphi=\varphi_{\theta}$ have been made
for the generator in the literature, see e.g.~Table
\ref{lambert:ArchCop:Example}, where Kendall's tau is expressed as a
function of $\theta$.
\begin{table}[t!]\centering
\caption{Examples of parametric Archimedean copulas.\label{lambert:ArchCop:Example}}
\medskip
\begin{tabular}{lcccc}
Family & $\theta$ range & Generator & Kendall's tau \\
\hline
Frank & $(-\infty,+\infty)/\{0\}$ & $-\log{\mathrm{e}^{-\theta u}-1\over \mathrm{e}^{-\theta}-1}$ & 
  $1 - {4\over \theta}\left(1-{1\over\theta}\int_0^\theta {t\over \mathrm{e}^t-1}dt\right)$\\
  Clayton & $[-1,+\infty)/\{0\}$ & $(u^{-\theta}-1)/\theta$ & $\theta/(\theta+2)$ \\
  Gumbel & $[-1,+\infty)$ & $(-\log u)^\theta$ & $(\theta-1)/\theta$ \\
\hline
\end{tabular}
\end{table}
For a given choice of the copula family and independent pairs
$\{(u_i,v_i):i=1,\ldots,n\}$ of data with uniform margins, one can
estimate $\theta$ using e.g.~the maximum likelihood principle. One can
show that the likelihood is
\begin{eqnarray}
L(\theta|\D) &=& \prod_{i=1}^n {\partial^2 \over \partial u \partial v}C_\theta(u_i,v_i)
= -\prod_{i=1}^n
{\varphi^{\prime\prime}_\theta(C_i){\varphi^{\prime}_\theta}(u_i)
\varphi^{\prime}_\theta(v_i)
  \over \left({\varphi^{\prime}_\theta(C_i)}\right)^3
} \label{Likelihood:Eqn}
\end{eqnarray}
where $$C_i=\varphi_\theta^{-1}\left({\varphi_\theta(u_i)+\varphi_\theta(v_i)}\right)$$
and $\D$ generically denotes the available data.

\subsection{The spline approximation}
A (spline based) nonparametric estimate of $\varphi(\cdot)$ was first
proposed in \citet{Lambert:2007vo}.
Here, we present another formulation with superior properties and
embedding the proposal made by \citet{Vandenhende:2005ug} where the
function $g(\cdot)$ in (\ref{lambert:NonparametricGenerator}) was
piecewise linear. It can be written as
\begin{eqnarray}
  \varphi_{\vec\theta}(u) = \exp\left\{-g_{\vec\theta}(S(u))\right\} \label{lambert:NonparametricGenerator}
\end{eqnarray}
where $S(u)=-\log(-\log(u))$ is the quantile function of the extreme
value distribution and 
\begin{align}
{d\over ds}g_{\vec\theta}(s) &= \sum_{k=1}^K b_{k}(s)(1+\theta^2_{k})
\nonumber \\
& = 1+ \sum_{k=1}^K b_{k}(s) \theta^2_{k} \label{derivativeOfG:Eq}
\end{align}
is a linear combination of $K$ cubic B-splines associated to
equidistant knots on $(S(\epsilon),S(1-\epsilon))$ for a small
quantity $\epsilon$ ($=10^{-6}$, say). One can show that such a
formulation ensures that (\ref{lambert:NonparametricGenerator})
provides a valid generator for any
$\vec\theta=(\theta_1,\ldots,\theta_K)'$ in $\rit^K$. In the special
case where $\theta_k=\theta$ for all $k$, the generator is
$\varphi(\cdot)=\left(-\log(\cdot)\right)^\zeta$, i.e.~that of
a Gumbel copula with dependence parameter
$\zeta=1+\theta^2$. When, in addition, $\theta=0$, one obtains the
independence copula. 

\subsection{Inference} \label{CopulaSpline:Inference:Sect} 

Given $K$, the inferential problem ends up to the selection of the
parameters $\vec{\theta}$ defining the spline coefficients. We suggest
to follow the proposal made by \citet{EilersP:flex:96} by taking a
large number of equidistant knots and to counterbalance the introduced
flexibility by penalizing changes in $r$th order differences of the
spline coefficients. Then, the penalized log-likelihood (see
\ref{Appendix:1} for computational details) is
$$l_{pen}(\vec\theta|\D) = \log L(\vec\theta|\D) - {\kappa \over 2} \vec{\theta}' P\vec{\theta}$$
where $P=D_r'D_r$ and $D_r$ is the $(K-r)\times K$ matrix yielding
$r$th order differences of the splines coefficients when applied on
$\vec\theta$.

The penalty parameter $\kappa$ can be selected using cross validation
or an information criterion. A possible translation in Bayesian terms
\citep{LangS:Bayes:04} takes as prior for $\vec\theta$
\begin{eqnarray*}
 p(\vec\theta|\kappa) \propto \kappa^{\rho(P)/2}\exp\left(-{\kappa\over 2}
   \vec\theta'P\vec\theta \right) 
\end{eqnarray*}
where $\rho(P)$ denotes the rank of $P$. A gamma distribution is a
possible choice for the prior of the penalty parameter,
$\kappa\sim{\cal G}(a,b)$. The marginal posterior distribution for
$\theta$ \citep{Jullion:2007vu} is obtained by integrating out
$\kappa$ from the joint posterior for $(\vec\theta,\kappa)$:
\begin{eqnarray}
p(\vec\theta|\D) \propto \int_0^{+\infty} L(\vec\theta|\D)\, p(\vec\theta|\kappa)
\, p(\kappa)\, d\kappa
\propto {L(\vec\theta|\D) \over \left(b +{1\over
      2}\vec\theta'P\vec\theta \right)^{a+{\rho(P)\over 2}}}
\label{MarginalPosterior:Eq}
\end{eqnarray}
That expression reveals that it is equivalent to assuming the
following independent Student priors for the differences
$D_r\vec\theta$ of the splines coefficients:
\begin{eqnarray}
(D_r\vec\theta |\D) \sim t_{\nu=2a}\left({0,{b\over a}I_{\rho(P)}}\right),
\label{StudentPriorEq}
\end{eqnarray}
where $t_{\nu}(\mu,\Sigma)$ is the multivariate Student-t distribution
with $\nu$ degrees of freedom, mean $\mu$ and variance-covariance
matrix ${\nu\over \nu-2}\Sigma$ when these two moments
exist. Therefore, taking $a=1$ and a 'large' value for $b$ ($1$, say)
leaves a priori a lot of freedom to the spline coefficients with a
regularization of their $r$th order differences towards 0. The
pertinence of that recommendation will be confirmed by the simulation
study in Section \ref{CopulaSpline:SimulationStudy:Sect}.

A possible point estimate for $\vec\theta$, and, hence, for the copula
generator, can be obtained by maximizing
(\ref{MarginalPosterior:Eq}), yielding the maximum posterior
probability (MAP) estimate $\vec\hat\theta$.  A sample
$\left\{\vec\theta^{(m)}:m=1,\ldots,M\right\}$ from the joint
posterior for $\vec\theta$ can also be obtained using an importance
sampler with vectors generated using the multivariate Student
distribution
$t_{K}\left(\hat{\vec\theta},(-H_{\hat{\vec\theta}})^{-1}\right)$
where $H_{\hat{\vec\theta}}$ denotes the Hessian matrix evaluated at
$\hat{\vec\theta}$.

\subsection{Simulation  study} \label{CopulaSpline:SimulationStudy:Sect}
A first simulation study was set up to evaluate the merits of the
spline model to estimate the generator of an Archimedean
copula. $S=500$ datasets $\{(u_{1i},u_{2i}):i=1,\ldots,n\}$ were
generated with $n=100$, 250, 500 or 2000 pairs with uniform margins
and an association structure characterized by a Clayton, a Frank or a
Gumbel copula with a dependence parameter corresponding to a Kendall's
tau equal to $0.15$, $0.30$ or $0.45$. The copula generator was
approximated using
(\ref{lambert:NonparametricGenerator}-\ref{derivativeOfG:Eq})
with $K=11$ knots, a 3rd order penalty, and a gamma prior for the
penalty coefficient, $\kappa\sim{\cal G}(a=1,b=1)$.

\begin{table}[t]
\centering
{\small
\begin{tabular}{rrrrrrrrrr}
& \multicolumn{9}{c}{Estimation of $\lambda(u)$ -- Clayton copula ($\tau=0.15$)}\\
  \cline{3-10}
 &      & \multicolumn{2}{c}{$n=100$} & \multicolumn{2}{c}{$n=250$} & \multicolumn{2}{c}{$n=500$} & \multicolumn{2}{c}{$n=2000$} \\
$u$ & $\lambda(u)$ & Bias & RMSE & Bias & RMSE & Bias & RMSE & Bias & RMSE \\ 
  \hline
  0.05 & -0.092 & -0.013 & 0.021 & -0.006 & 0.014 & -0.004 & 0.010 & -0.001 & 0.005 \\ 
  0.10 & -0.158 & -0.012 & 0.026 & -0.005 & 0.017 & -0.002 & 0.012 & -0.001 & 0.006 \\ 
  0.20 & -0.246 & -0.006 & 0.028 & -0.002 & 0.019 & 0.000 & 0.013 & -0.000 & 0.007 \\ 
  0.30 & -0.294 & -0.001 & 0.026 & 0.000 & 0.019 & 0.001 & 0.014 & -0.000 & 0.007 \\ 
  0.40 & -0.313 & 0.002 & 0.023 & 0.001 & 0.017 & 0.001 & 0.013 & -0.000 & 0.007 \\ 
  0.50 & -0.307 & 0.004 & 0.019 & 0.002 & 0.015 & 0.000 & 0.011 & -0.000 & 0.006 \\ 
  0.60 & -0.280 & 0.004 & 0.015 & 0.002 & 0.011 & 0.000 & 0.009 & 0.000 & 0.006 \\ 
  0.70 & -0.235 & 0.005 & 0.011 & 0.002 & 0.008 & 0.001 & 0.006 & 0.000 & 0.004 \\ 
  0.80 & -0.172 & 0.005 & 0.008 & 0.002 & 0.005 & 0.001 & 0.004 & 0.000 & 0.003 \\ 
  0.90 & -0.093 & 0.004 & 0.006 & 0.002 & 0.003 & 0.001 & 0.002 & 0.001 & 0.001 \\ 
  0.95 & -0.048 & 0.004 & 0.005 & 0.002 & 0.002 & 0.001 & 0.001 & 0.000 & 0.001 \\
   \hline
\end{tabular}
}
\wcaption{Estimation of $\lambda(u)=\varphi(u)/\varphi'(u)$ on a grid
  of values for $u$ using the spline estimator in
  (\ref{lambert:NonparametricGenerator}).  Biases and RMSE's were
  estimated using $S=500$ randomly generated datasets of size
  $n$.}{TabClayton15} 
\end{table}

\begin{table}[t]
\centering
{\small
\begin{tabular}{rrrrrrrrrr}
& \multicolumn{9}{c}{Estimation of $\lambda(u)$ -- Clayton copula ($\tau=0.30$)}\\
  \cline{3-10}
 &         & \multicolumn{2}{c}{$n=100$} & \multicolumn{2}{c}{$n=250$} & \multicolumn{2}{c}{$n=500$} & \multicolumn{2}{c}{$n=2000$} \\
$u$ & $\lambda(u)$ & Bias & RMSE & Bias & RMSE & Bias & RMSE & Bias & RMSE \\ 
  \hline
  0.05 & -0.054 & -0.007 & 0.016 & -0.004 & 0.009 & -0.002 & 0.006 & -0.000 & 0.003 \\ 
  0.10 & -0.100 & -0.007 & 0.020 & -0.002 & 0.011 & -0.001 & 0.008 & -0.001 & 0.005 \\ 
  0.20 & -0.175 & -0.003 & 0.022 & 0.003 & 0.014 & 0.000 & 0.010 & -0.000 & 0.005 \\ 
  0.30 & -0.225 & 0.000 & 0.024 & 0.005 & 0.018 & 0.001 & 0.013 & -0.000 & 0.006 \\ 
  0.40 & -0.254 & 0.003 & 0.024 & 0.005 & 0.019 & 0.001 & 0.014 & -0.000 & 0.007 \\ 
  0.50 & -0.261 & 0.005 & 0.023 & 0.003 & 0.018 & 0.001 & 0.013 & 0.000 & 0.006 \\ 
  0.60 & -0.248 & 0.005 & 0.021 & 0.001 & 0.016 & 0.001 & 0.012 & 0.001 & 0.006 \\ 
  0.70 & -0.215 & 0.005 & 0.017 & -0.000 & 0.011 & 0.001 & 0.009 & 0.001 & 0.004 \\ 
  0.80 & -0.162 & 0.005 & 0.012 & 0.000 & 0.006 & 0.001 & 0.006 & 0.000 & 0.003 \\ 
  0.90 & -0.091 & 0.005 & 0.008 & 0.001 & 0.003 & 0.002 & 0.003 & 0.000 & 0.002 \\ 
  0.95 & -0.048 & 0.004 & 0.006 & 0.002 & 0.003 & 0.001 & 0.002 & 0.000 & 0.001 \\
   \hline
\end{tabular}
}
\wcaption{Estimation of $\lambda(u)=\varphi(u)/\varphi'(u)$ on a grid
  of values for $u$ using the spline estimator in
  (\ref{lambert:NonparametricGenerator}).  Biases and RMSE's were
  estimated using $S=500$ randomly generated datasets of size
  $n$.}{TabClayton30} 
\end{table}

\begin{table}[t]
\centering
{\small
\begin{tabular}{rrrrrrrrrr}
& \multicolumn{9}{c}{Estimation of $\lambda(u)$ -- Clayton copula ($\tau=0.45$)}\\
  \cline{3-10}
 &         & \multicolumn{2}{c}{$n=100$} & \multicolumn{2}{c}{$n=250$} & \multicolumn{2}{c}{$n=500$} & \multicolumn{2}{c}{$n=2000$} \\
$u$ & $\lambda(u)$ & Bias & RMSE & Bias & RMSE & Bias & RMSE & Bias & RMSE \\ 
  \hline
  0.05 & -0.030 & -0.003 & 0.008 & -0.004 & 0.006 & 0.000 & 0.003 & 0.000 & 0.002 \\ 
  0.10 & -0.060 & -0.002 & 0.011 & -0.002 & 0.007 & 0.001 & 0.005 & 0.001 & 0.003 \\ 
  0.20 & -0.113 & 0.000 & 0.016 & 0.003 & 0.009 & 0.001 & 0.007 & 0.001 & 0.004 \\ 
  0.30 & -0.158 & 0.000 & 0.021 & 0.007 & 0.013 & -0.001 & 0.010 & 0.000 & 0.005 \\ 
  0.40 & -0.190 & 0.001 & 0.023 & 0.008 & 0.016 & -0.002 & 0.011 & -0.001 & 0.006 \\ 
  0.50 & -0.207 & 0.003 & 0.022 & 0.006 & 0.016 & -0.001 & 0.011 & -0.000 & 0.006 \\ 
  0.60 & -0.208 & 0.003 & 0.021 & 0.002 & 0.015 & -0.001 & 0.012 & 0.000 & 0.006 \\ 
  0.70 & -0.189 & 0.003 & 0.018 & -0.001 & 0.012 & 0.000 & 0.010 & 0.000 & 0.005 \\ 
  0.80 & -0.150 & 0.004 & 0.013 & -0.001 & 0.008 & 0.001 & 0.007 & 0.000 & 0.004 \\ 
  0.90 & -0.087 & 0.005 & 0.009 & 0.001 & 0.004 & 0.002 & 0.004 & 0.000 & 0.002 \\ 
  0.95 & -0.047 & 0.005 & 0.007 & 0.001 & 0.002 & 0.001 & 0.002 & 0.000 & 0.001 \\ 
   \hline
\end{tabular}
}
\wcaption{Estimation of $\lambda(u)=\varphi(u)/\varphi'(u)$ on a grid
  of values for $u$ using the spline estimator in
  (\ref{lambert:NonparametricGenerator}).  Biases and RMSE's were
  estimated using $S=500$ randomly generated datasets of size
  $n$.}{TabClayton45} 
\end{table}

\begin{table}[t]
\centering
{\small
\begin{tabular}{rrrrrrrrrr}
& \multicolumn{9}{c}{Estimation of $\lambda(u)$ -- Frank copula ($\tau=0.15$)}\\
  \cline{3-10}
 &      & \multicolumn{2}{c}{$n=100$} & \multicolumn{2}{c}{$n=250$} & \multicolumn{2}{c}{$n=500$} & \multicolumn{2}{c}{$n=2000$} \\
$u$ & $\lambda(u)$ & Bias & RMSE & Bias & RMSE & Bias & RMSE & Bias & RMSE \\ 
  \hline
  0.05 & -0.125 & 0.005 & 0.016 & 0.004 & 0.011 & 0.003 & 0.008 & 0.001 & 0.004 \\ 
  0.10 & -0.189 & 0.000 & 0.021 & 0.000 & 0.015 & 0.001 & 0.010 & -0.000 & 0.005 \\ 
  0.20 & -0.261 & -0.008 & 0.028 & -0.007 & 0.020 & -0.004 & 0.013 & -0.002 & 0.007 \\ 
  0.30 & -0.295 & -0.013 & 0.030 & -0.010 & 0.021 & -0.006 & 0.014 & -0.002 & 0.007 \\ 
  0.40 & -0.303 & -0.013 & 0.028 & -0.009 & 0.020 & -0.005 & 0.014 & -0.002 & 0.007 \\ 
  0.50 & -0.293 & -0.010 & 0.025 & -0.007 & 0.018 & -0.003 & 0.012 & -0.001 & 0.006 \\ 
  0.60 & -0.266 & -0.006 & 0.020 & -0.004 & 0.015 & -0.001 & 0.010 & -0.000 & 0.006 \\ 
  0.70 & -0.223 & -0.002 & 0.014 & -0.001 & 0.011 & 0.001 & 0.008 & 0.000 & 0.005 \\ 
  0.80 & -0.165 & 0.002 & 0.009 & 0.001 & 0.007 & 0.001 & 0.005 & -0.000 & 0.003 \\ 
  0.90 & -0.091 & 0.003 & 0.006 & 0.002 & 0.004 & 0.001 & 0.003 & 0.000 & 0.001 \\ 
  0.95 & -0.048 & 0.003 & 0.004 & 0.002 & 0.003 & 0.001 & 0.001 & 0.000 & 0.001 \\ 
   \hline
\end{tabular}
}
\wcaption{Estimation of $\lambda(u)=\varphi(u)/\varphi'(u)$ on a grid
  of values for $u$ using the spline estimator in
  (\ref{lambert:NonparametricGenerator}).  Biases and RMSE's were
  estimated using $S=500$ randomly generated datasets of size
  $n$.}{TabFrank15} 
\end{table}

\begin{table}[t]
\centering
{\small
\begin{tabular}{rrrrrrrrrr}
& \multicolumn{9}{c}{Estimation of $\lambda(u)$ -- Frank copula ($\tau=0.30$)}\\
  \cline{3-10}
 &      & \multicolumn{2}{c}{$n=100$} & \multicolumn{2}{c}{$n=250$} & \multicolumn{2}{c}{$n=500$} & \multicolumn{2}{c}{$n=2000$} \\
$u$ & $\lambda(u)$ & Bias & RMSE & Bias & RMSE & Bias & RMSE & Bias & RMSE \\ 
  \hline
  0.05 & -0.105 & 0.009 & 0.017 & 0.007 & 0.012 & 0.004 & 0.008 & -0.000 & 0.004 \\ 
  0.10 & -0.153 & 0.004 & 0.021 & 0.003 & 0.013 & 0.001 & 0.009 & -0.000 & 0.005 \\ 
  0.20 & -0.207 & -0.008 & 0.026 & -0.005 & 0.016 & -0.005 & 0.012 & -0.000 & 0.006 \\ 
  0.30 & -0.232 & -0.014 & 0.030 & -0.009 & 0.017 & -0.006 & 0.013 & -0.000 & 0.006 \\ 
  0.40 & -0.241 & -0.015 & 0.029 & -0.008 & 0.017 & -0.005 & 0.013 & -0.000 & 0.006 \\ 
  0.50 & -0.237 & -0.012 & 0.027 & -0.005 & 0.016 & -0.002 & 0.012 & -0.000 & 0.006 \\ 
  0.60 & -0.220 & -0.007 & 0.022 & -0.002 & 0.014 & 0.001 & 0.010 & -0.000 & 0.006 \\ 
  0.70 & -0.192 & -0.001 & 0.018 & 0.001 & 0.012 & 0.002 & 0.009 & 0.000 & 0.004 \\ 
  0.80 & -0.148 & 0.002 & 0.012 & 0.002 & 0.009 & 0.001 & 0.006 & -0.000 & 0.004 \\ 
  0.90 & -0.086 & 0.003 & 0.007 & 0.002 & 0.005 & 0.001 & 0.003 & 0.000 & 0.002 \\ 
  0.95 & -0.046 & 0.003 & 0.005 & 0.001 & 0.003 & 0.001 & 0.002 & 0.000 & 0.001 \\ 
   \hline
\end{tabular}
}
\wcaption{Estimation of $\lambda(u)=\varphi(u)/\varphi'(u)$ on a grid
  of values for $u$ using the spline estimator in
  (\ref{lambert:NonparametricGenerator}).  Biases and RMSE's were
  estimated using $S=500$ randomly generated datasets of size
  $n$.}{TabFrank30} 
\end{table}

\begin{table}[t]
\centering
{\small
\begin{tabular}{rrrrrrrrrr}
& \multicolumn{9}{c}{Estimation of $\lambda(u)$ -- Frank copula ($\tau=0.45$)}\\
  \cline{3-10}
 &      & \multicolumn{2}{c}{$n=100$} & \multicolumn{2}{c}{$n=250$} & \multicolumn{2}{c}{$n=500$} & \multicolumn{2}{c}{$n=2000$} \\
$u$ & $\lambda(u)$ & Bias & RMSE & Bias & RMSE & Bias & RMSE & Bias & RMSE \\ 
  \hline
  0.05 & -0.086 & 0.012 & 0.018 & 0.009 & 0.014 & 0.003 & 0.007 & 0.001 & 0.003 \\ 
  0.10 & -0.121 & 0.008 & 0.020 & 0.005 & 0.013 & 0.002 & 0.009 & 0.001 & 0.004 \\ 
  0.20 & -0.157 & -0.004 & 0.023 & -0.004 & 0.014 & -0.001 & 0.010 & 0.000 & 0.005 \\ 
  0.30 & -0.174 & -0.010 & 0.025 & -0.009 & 0.016 & -0.002 & 0.010 & -0.001 & 0.005 \\ 
  0.40 & -0.180 & -0.011 & 0.024 & -0.010 & 0.016 & -0.003 & 0.009 & -0.001 & 0.005 \\ 
  0.50 & -0.179 & -0.010 & 0.024 & -0.007 & 0.014 & -0.002 & 0.010 & -0.001 & 0.005 \\ 
  0.60 & -0.171 & -0.005 & 0.022 & -0.002 & 0.012 & -0.001 & 0.011 & -0.000 & 0.005 \\ 
  0.70 & -0.155 & 0.000 & 0.018 & 0.002 & 0.011 & 0.001 & 0.009 & 0.001 & 0.004 \\ 
  0.80 & -0.127 & 0.003 & 0.014 & 0.003 & 0.009 & -0.000 & 0.008 & -0.000 & 0.004 \\ 
  0.90 & -0.079 & 0.005 & 0.010 & 0.002 & 0.006 & 0.001 & 0.004 & 0.000 & 0.002 \\ 
  0.95 & -0.044 & 0.004 & 0.006 & 0.001 & 0.003 & 0.001 & 0.003 & 0.000 & 0.001 \\ 
   \hline
\end{tabular}
}
\wcaption{Estimation of $\lambda(u)=\varphi(u)/\varphi'(u)$ on a grid
  of values for $u$ using the spline estimator in
  (\ref{lambert:NonparametricGenerator}).  Biases and RMSE's were
  estimated using $S=500$ randomly generated datasets of size
  $n$.}{TabFrank45} 
\end{table}

\begin{table}[t]
\centering
{\small
\begin{tabular}{rrrrrrrrrr}
& \multicolumn{9}{c}{Estimation of $\lambda(u)$ -- Gumbel copula ($\tau=0.15$)}\\
  \cline{3-10}
 &      & \multicolumn{2}{c}{$n=100$} & \multicolumn{2}{c}{$n=250$} & \multicolumn{2}{c}{$n=500$} & \multicolumn{2}{c}{$n=2000$} \\
$u$ & $\lambda(u)$ & Bias & RMSE & Bias & RMSE & Bias & RMSE & Bias & RMSE \\ 
  \hline
  0.05 & -0.127 & 0.002 & 0.013 & 0.001 & 0.011 & -0.001 & 0.009 & -0.001 & 0.005 \\ 
  0.10 & -0.196 & 0.001 & 0.018 & -0.001 & 0.015 & -0.002 & 0.012 & -0.001 & 0.006 \\ 
  0.20 & -0.274 & -0.004 & 0.022 & -0.005 & 0.018 & -0.003 & 0.014 & -0.001 & 0.007 \\ 
  0.30 & -0.307 & -0.006 & 0.024 & -0.006 & 0.019 & -0.004 & 0.015 & -0.000 & 0.007 \\ 
  0.40 & -0.312 & -0.008 & 0.024 & -0.006 & 0.019 & -0.003 & 0.014 & 0.000 & 0.007 \\ 
  0.50 & -0.295 & -0.008 & 0.022 & -0.006 & 0.018 & -0.003 & 0.013 & -0.000 & 0.006 \\ 
  0.60 & -0.261 & -0.007 & 0.020 & -0.005 & 0.016 & -0.002 & 0.011 & -0.000 & 0.005 \\ 
  0.70 & -0.212 & -0.005 & 0.017 & -0.003 & 0.013 & -0.001 & 0.009 & -0.000 & 0.004 \\ 
  0.80 & -0.152 & -0.003 & 0.012 & -0.003 & 0.010 & -0.001 & 0.007 & -0.000 & 0.003 \\ 
  0.90 & -0.081 & -0.001 & 0.007 & -0.001 & 0.006 & -0.001 & 0.004 & -0.001 & 0.002 \\ 
  0.95 & -0.041 & 0.001 & 0.005 & -0.000 & 0.004 & -0.000 & 0.003 & -0.001 & 0.002 \\ 
   \hline
\end{tabular}
}
\wcaption{Estimation of $\lambda(u)=\varphi(u)/\varphi'(u)$ on a grid
  of values for $u$ using the spline estimator in
  (\ref{lambert:NonparametricGenerator}).  Biases and RMSE's were
  estimated using $S=500$ randomly generated datasets of size
  $n$.}{TabGumbel15} 
\end{table}

\begin{table}[t]
\centering
{\small
\begin{tabular}{rrrrrrrrrr}
& \multicolumn{9}{c}{Estimation of $\lambda(u)$ -- Gumbel copula ($\tau=0.30$)}\\
  \cline{3-10}
 &      & \multicolumn{2}{c}{$n=100$} & \multicolumn{2}{c}{$n=250$} & \multicolumn{2}{c}{$n=500$} & \multicolumn{2}{c}{$n=2000$} \\
$u$ & $\lambda(u)$ & Bias & RMSE & Bias & RMSE & Bias & RMSE & Bias & RMSE \\ 
  \hline
  0.05 & -0.105 & 0.001 & 0.017 & 0.000 & 0.011 & -0.001 & 0.008 & -0.001 & 0.005 \\ 
  0.10 & -0.161 & -0.002 & 0.022 & -0.001 & 0.015 & -0.002 & 0.011 & -0.001 & 0.006 \\ 
  0.20 & -0.225 & -0.006 & 0.027 & -0.002 & 0.018 & -0.002 & 0.013 & -0.001 & 0.007 \\ 
  0.30 & -0.253 & -0.007 & 0.029 & -0.002 & 0.018 & -0.001 & 0.013 & -0.000 & 0.007 \\ 
  0.40 & -0.257 & -0.007 & 0.029 & -0.001 & 0.017 & -0.000 & 0.012 & 0.000 & 0.006 \\ 
  0.50 & -0.243 & -0.007 & 0.028 & -0.001 & 0.015 & 0.000 & 0.011 & -0.000 & 0.006 \\ 
  0.60 & -0.215 & -0.006 & 0.026 & -0.001 & 0.013 & 0.000 & 0.010 & -0.000 & 0.005 \\ 
  0.70 & -0.175 & -0.005 & 0.022 & -0.001 & 0.011 & 0.000 & 0.008 & -0.000 & 0.004 \\ 
  0.80 & -0.125 & -0.005 & 0.017 & -0.002 & 0.010 & -0.000 & 0.007 & -0.000 & 0.004 \\ 
  0.90 & -0.066 & -0.003 & 0.010 & -0.000 & 0.006 & -0.001 & 0.005 & -0.000 & 0.003 \\ 
  0.95 & -0.034 & -0.000 & 0.006 & -0.000 & 0.004 & -0.001 & 0.003 & 0.000 & 0.002 \\ 
   \hline
\end{tabular}
}
\wcaption{Estimation of $\lambda(u)=\varphi(u)/\varphi'(u)$ on a grid
  of values for $u$ using the spline estimator in
  (\ref{lambert:NonparametricGenerator}).  Biases and RMSE's were
  estimated using $S=500$ randomly generated datasets of size
  $n$.}{TabGumbel30} 
\end{table}

\begin{table}[t]
\centering
{\small
\begin{tabular}{rrrrrrrrrr}
& \multicolumn{9}{c}{Estimation of $\lambda(u)$ -- Gumbel copula ($\tau=0.45$)}\\
  \cline{3-10}
 &      & \multicolumn{2}{c}{$n=100$} & \multicolumn{2}{c}{$n=250$} & \multicolumn{2}{c}{$n=500$} & \multicolumn{2}{c}{$n=2000$} \\
$u$ & $\lambda(u)$ & Bias & RMSE & Bias & RMSE & Bias & RMSE & Bias & RMSE \\ 
  \hline
  0.05 & -0.082 & 0.003 & 0.014 & -0.001 & 0.009 & -0.001 & 0.008 & -0.000 & 0.004 \\ 
  0.10 & -0.127 & 0.001 & 0.019 & -0.002 & 0.011 & -0.001 & 0.009 & -0.000 & 0.005 \\ 
  0.20 & -0.177 & -0.003 & 0.023 & -0.002 & 0.013 & 0.000 & 0.011 & 0.000 & 0.006 \\ 
  0.30 & -0.199 & -0.004 & 0.024 & -0.001 & 0.013 & 0.001 & 0.011 & 0.000 & 0.006 \\ 
  0.40 & -0.202 & -0.004 & 0.025 & -0.001 & 0.012 & 0.000 & 0.010 & -0.000 & 0.005 \\ 
  0.50 & -0.191 & -0.005 & 0.026 & -0.001 & 0.012 & -0.001 & 0.011 & -0.001 & 0.005 \\ 
  0.60 & -0.169 & -0.004 & 0.023 & -0.001 & 0.011 & -0.002 & 0.010 & -0.001 & 0.005 \\ 
  0.70 & -0.137 & -0.003 & 0.019 & -0.000 & 0.009 & -0.001 & 0.008 & -0.001 & 0.004 \\ 
  0.80 & -0.098 & -0.003 & 0.016 & -0.000 & 0.007 & -0.001 & 0.007 & -0.000 & 0.004 \\ 
  0.90 & -0.052 & -0.000 & 0.010 & -0.000 & 0.005 & 0.000 & 0.005 & -0.000 & 0.002 \\ 
  0.95 & -0.027 & 0.001 & 0.006 & -0.000 & 0.004 & -0.000 & 0.003 & -0.000 & 0.001 \\ 
   \hline
\end{tabular}
}
\wcaption{Estimation of $\lambda(u)=\varphi(u)/\varphi'(u)$ on a grid
  of values for $u$ using the spline estimator in
  (\ref{lambert:NonparametricGenerator}).  Biases and RMSE's were
  estimated using $S=500$ randomly generated datasets of size
  $n$.}{TabGumbel45} 
\end{table}

\begin{table}[t]
\centering
\begin{tabular}{lccccc}
        &        & \multicolumn{4}{c}{Sample size $n$}\\
\cline{3-6}        
Copula  & $\tau$ & {$100$} & {$250$} & {$500$} & {$2000$} \\
\hline
Clayton & $0.15$ & 0.0147 & 0.0110 & 0.0081 & 0.0044\\
        & $0.30$ & 0.0153 & 0.0122 & 0.0087 & 0.0043\\
        & $0.45$ & 0.0149 & 0.0106 & 0.0075 & 0.0037\\
Frank   & $0.15$ & 0.0158 & 0.0105 & 0.0075 & 0.0043\\
        & $0.30$ & 0.0172 & 0.0107 & 0.0081 & 0.0042\\
        & $0.45$ & 0.0187 & 0.0116 & 0.0078 & 0.0038\\
Gumbel  & $0.15$ & 0.0153 & 0.0118 & 0.0078 & 0.0046\\
        & $0.30$ & 0.0186 & 0.0109 & 0.0083 & 0.0046\\
        & $0.45$ & 0.0182 & 0.0073 & 0.0079 & 0.0040\\
   \hline
\end{tabular}
\wcaption{Estimation of $\lambda(u)=\varphi(u)/\varphi'(u)$ over
  $(0,1)$: RMISE of the spline estimator in
  (\ref{lambert:NonparametricGenerator}) estimated using $S=500$
  randomly generated datasets of size $n$.}{RMISEtab} 
\end{table}

\begin{table}[t]
\centering
\begin{tabular}{lrrrrrrrrrr}
 \cline{3-11}
& &\multicolumn{3}{c}{$\tau=0.45$} & \multicolumn{3}{c}{$\tau=0.30$} &
\multicolumn{3}{c}{$\tau=0.15$}\\
  \cline{3-11}
& & \multicolumn{9}{c}{Nominal coverage} \\
 \cline{3-11}
& $n$ & 0.80 & 0.90 & 0.95 & 0.80 & 0.90 & 0.95 & 0.80 & 0.90 & 0.95 \\ 
 \cline{2-11}
Clayton&  100 & 0.83 & 0.91 & 0.94 & 0.83 & 0.92 & 0.95 & 0.87 & 0.95 & 0.97\\ 
&  250 & 0.84 & 0.93 & 0.97 & 0.84 & 0.92 & 0.96 & 0.86 & 0.94 & 0.97\\ 
 & 500 & 0.82 & 0.91 & 0.96 & 0.82 & 0.91 & 0.96 & 0.87 & 0.95 & 0.98\\ 
&2000 & 0.81 & 0.91 & 0.96 & 0.82 & 0.92 & 0.97 & 0.84 & 0.93 & 0.96\\
&\\
Frank &  100& 0.80 & 0.91 & 0.95 & 0.85 & 0.93 & 0.96 & 0.87 & 0.96 & 0.99 \\ 
&  250& 0.82 & 0.91 & 0.96 & 0.86 & 0.94 & 0.97 & 0.86 & 0.94 & 0.97 \\ 
&  500& 0.82 & 0.91 & 0.96 & 0.86 & 0.94 & 0.97 & 0.87 & 0.95 & 0.98 \\ 
&2000& 0.82 & 0.91 & 0.96 & 0.84 & 0.92 & 0.96  & 0.85 & 0.94 & 0.97\\ 
&\\
Gumbel & 100  & 0.84 & 0.93 & 0.96 & 0.82 & 0.90 & 0.94& 0.89 & 0.95 & 0.98\\ 
& 250  & 0.90 & 0.96 & 0.98 & 0.86 & 0.95 & 0.98& 0.83 & 0.93 & 0.97\\ 
& 500  & 0.83 & 0.92 & 0.96 & 0.86 & 0.94 & 0.97& 0.84 & 0.92 & 0.96\\ 
& 2000& 0.81 & 0.91 & 0.96 & 0.82 & 0.92 & 0.96& 0.83 & 0.92 & 0.97\\
 \cline{2-11}
\end{tabular}
\wcaption{Mean coverages of pointwise credible intervals for
  $\lambda(u)$ estimated using $S=500$ randomly generated datasets of
  size $n$ under the Clayton, Frank or Gumbel copula with different
  Kendall's $\tau$.}{TabCoverages}  
\end{table}

The posterior mode (MAP) $\hat\theta$ for the spline coefficients was
first estimated by maximizing (\ref{MarginalPosterior:Eq})
w.r.t.~$\theta$.  Then, a sample
$\left\{\theta^{(m)}:m=1,\ldots,M\right\}$ from the joint posterior
for $\theta$ was obtained using an importance sampler, see Section
\ref{CopulaSpline:Inference:Sect} for details.  The posterior mean
$\tilde\lambda(u)$ and pointwise credible intervals for
$\lambda(u)=\varphi(u)/\varphi'(u)$ were estimated using the induced
sample $\{\lambda(u|\theta^{(m)}):m=1,\ldots,M\}$ and the associated
importance weights \citep{Chen:1999}.  The bias and the RMSE of
$\tilde\lambda(u)$ for a grid of values for $u$ in $(0.05,.95)$ are
reported in Tables \ref{TabClayton15} to \ref{TabGumbel45}. It
suggests that the biases (if any) are very small, even with small
sample sizes. Root mean squared errors (RMSE) decrease with sample
size, with empirical results suggesting that they are proportional to
$n^{-1/2}$. The root mean integrated squared errors (RMISE),
$$RMISE = \left(\int_0^1(\lambda(u)-\tilde\lambda(u))^2 du \right)^{1/2}$$
were also computed as a summary value of the quality of the generator
estimator, see Table \ref{RMISEtab}.

The (mean) coverages of the pointwise credible intervals (computed
from a grid of 19 equidistant values for $u$ between 0.05 and 0.95)
for $\lambda(u)$
are reported in Table \ref{TabCoverages}. They tend to be slightly
larger than their nominal values with better performances obtained for
larger underlying Kendall's tau.

The number $K$ of B-splines in the basis should be large enough to
ensure sufficient flexibility to the approximation of the copula
generator. That flexibility is counter-balanced by the penalty part,
resulting in the Student prior for $r$th order differences of the
spline coefficients, see (\ref{StudentPriorEq}). The simulation
results suggest that $K=11$ is indeed sufficient to
have an estimator with low bias for the generator.

Besides the excellent statistical properties of the generator
estimator (including a low bias and a good agreement between the
nominal and the effective coverages of the computed credible regions),
the estimated functions turn to be smooth whatever the sample size.

\section{Flexible conditional Archimedean copula families}

The flexible form presented in Section \ref{CopulaSpline:Sect} for the
generator can be generalized by letting the spline coefficients change
smoothly with a continuous covariate $X$. Extending the preceding
definitions, we propose to take as conditional generator for the
Archimedean copula when $X=x$,
\begin{eqnarray}
  \varphi_{\Theta}(u|x) = \exp\left\{-g_{\Theta}(S(u)|x)\right\} \label{lambert:CondNonparametricGenerator}
\end{eqnarray}
where 
\begin{align}
{d\over ds}g_{\Theta}(s|x) & = \sum_{k=1}^K 
b_{k}(s) \left(1+ \theta^2_{k}(x) \right) 
\label{derivativeOfCondG:Eq1} \\ 
& = \sum_{k=1}^K 
b_{k}(s) \left\{1+ 
\left(\sum_{\ell=1}^{K^*} b^*_{\ell}(x) \theta_{k\ell}\right)^2\right\}
 \label{derivativeOfCondG:Eq2}
\end{align}
for a B-spline basis $\{b_{\ell}^{*}(\cdot)\}_{\ell=1}^{K^*}$ on the
domain $\cal X$ of the covariate values, a $K\times K^*$ matrix
$\Theta=(\theta_{kl})$ of spline coefficients and
$S(u)=-\log(-\log(u))$.  It directly affects Kendall's tau and relates
it to the covariate. Indeed, one can show that
\begin{align*}
  & \tau_{\Theta}(x) = 1 + 4\int_0^1 {\lambda_{\Theta}(u|x)} du
\end{align*}
where
\begin{align*}
 & \lambda_{\Theta}(u|x) = 
   {\varphi_{\Theta}(u|x) \over \varphi'_{\Theta}(u|x)}
  = -\left\{g'_{\Theta}(S(u)|x)\,S'(u)\right\}^{-1} \\
\end{align*}
This approximation to the conditional copula generator shares the
qualities of the spline approximation to the generator proposed in
Section \ref{CopulaSpline:Sect}. The price to pay is in the number of
spline parameters: $K\times K^*$ instead of $K$ in the unconditional
case. It motivates the two extra approximations proposed below in
Sections \ref{AdditiveCondSplineFamily:Sect} and
\ref{FlexPowerFamily:Sect}.

Smoothness can be forced using penalties on the $r_1$th order
differences in the $\theta_{kl}$'s for a given $\ell$ and on their
$r_2$th order differences for a given $k$, yielding the following
quadratic form for the penalty
$$ \mathrm{pen}(\Theta|\kappa_1,\kappa_2) = \mathrm{vec}(\Theta)'\, 
(\kappa_1 I_{K^*}\otimes P_1 + \kappa_2 P_2\otimes I_{K})\,
\mathrm{vec}(\Theta)
$$
where $P_1=D'_{r_1}D_{r_1}$ and $P_2=D'_{r_2}D_{r_2}$ denote the
penalty matrices of sizes $K$ and $K^*$, respectively, and $I_r$ the
identity matrix of size $r$.  The first term in the penalty induces
smoothness on $g_{\Theta}(s|x)$ in the $s-$dimension, while in the
second case, smoothness is obtained in the $x-$direction.

\subsection{The additive conditional spline Archimedean copula family} 
\label{AdditiveCondSplineFamily:Sect}

One can reduce the number of spline parameters to $(K+K^*-1)$ free
parameters using the additive form (\ref{derivativeOfCondG3:Eq}) for
$\theta_k(x)$ in (\ref{derivativeOfCondG:Eq1}):
\begin{eqnarray}
  \varphi_{\gamma,\beta}(u|x) = \exp\left\{-g_{\gamma,\beta}(S(u)|x)\right\} \label{lambert:CondNonparametricGenerator}
\end{eqnarray}
where 
\begin{align}
&{d\over ds}g_{\gamma,\beta}(s|x) = \sum_{k=1}^K b_{k}(s)\left(1+\theta ^2_{k}(x)\right)
\nonumber \\
&\theta_k(x) = \gamma_k + \beta(x)
 = \gamma_k + \sum_{\ell=1}^{K^*} b^*_{\ell}(x)\beta_\ell,  \label{derivativeOfCondG3:Eq}
\end{align}
for a B-spline basis $\{b_{\ell}^{*}(\cdot)\}_{\ell=1}^{K^*}$ on the
domain $\cal X$ of the covariate values, an identification constraint
$\sum_\ell \beta_\ell=0$ (say) and $S(u)=-\log(-\log(u))$.  Like in
the general case, one can show that the conditional Kendall's tau is
given by
\begin{align}
  & \tau_{\gamma,\beta}(x) = 1 + 4\int_0^1
  {\lambda_{\gamma,\beta}(u|x)} du
\label{CondKendallTau:PowerAlt:Eq}
\end{align}
where
\begin{align*}
 & \lambda_{\gamma,\beta}(u|x) = 
   {\varphi_{\gamma,\beta}(u|x) \over \varphi'_{\gamma,\beta}(u|x)}
  = -\left\{g'_{\gamma,\beta}(S(u)|x)\,S'(u)\right\}^{-1} \\
\end{align*}
It can be further extended to settings with multiple covariates. If
$\vec{x}=(x_1,\ldots,x_p)'$ denote $p$ continuous covariates, we
suggest to generalize (\ref{derivativeOfCondG3:Eq}) to
\begin{align}
&{d\over ds}g_{\gamma,\beta}(s|\vec{x}) = \sum_{k=1}^K
b_{k}(s)\left(1+\theta ^2_{k}(\vec{x})\right) 
\nonumber \\
&\theta_k(\vec{x}) = \gamma_k + \sum_{j=1}^p\beta_j(x_j)
 = \gamma_k + \sum_{j=1}^p \sum_{\ell=1}^{K^*} b^*_{\ell}(x_j)\beta_{j\ell},  \label{derivativeOfCondG4:Eq}
\end{align}
with identification constraints $\sum_\ell \beta_{j\ell}=0$ for all
$j$. That model contains $K+p(K^*-1)$ free parameters for the
conditional copula. 

\subsubsection{Inference} \label{PowerAltFamily:Inference:Sec}
We shall focus on the single covariate case, the extension to the
additive setting being straightforward. The likelihood is given by
\begin{align}
L(\vec\gamma,\vec\beta|\D) &=
 \prod_{i=1}^n {\partial^2 \over \partial u \partial v}
C_{\vec\gamma,\vec\beta}(u_i,v_i|x_i)
= -\prod_{i=1}^n
{\varphi^{\prime\prime}_{\vec\gamma,\vec\beta}(C_i|x_i)
\, \varphi^{\prime}_{\vec\gamma,\vec\beta}(u_i|x_i)
\, \varphi^{\prime}_{\vec\gamma,\vec\beta}(v_i|x_i)
\over 
 \left({\varphi^{\prime}_{\vec\gamma,\vec\beta}(C_i|x_i)}\right)^3
}, \label{Likelihood:Eqn2}
\end{align}
see (\ref{Likelihood:computation:Eq}) for computational details.
Smoothness is encouraged on $\varphi_{\gamma,\beta}(u|x)$ in the $u-$
and $x-$scales by penalizing changes in 2nd or 3rd order differences,
$D_\gamma \vec{\gamma}$ and $D_\beta \vec{\beta}$, of the spline
coefficients $\vec{\gamma}$ and $\vec{\beta}$ \citep{EilersP:flex:96}. The resulting
penalties,
\begin{align*}
 & \mathrm{pen}(\vec{\gamma}|\kappa_\gamma) 
= -{\kappa_\gamma \over 2} \, \vec{\gamma}' P_\gamma  \vec{\gamma}
\text{~~with~~} P_\gamma = D'_\gamma D_\gamma, \\
 & \mathrm{pen}(\vec{\beta}|\kappa_\beta) 
= -{\kappa_\beta \over 2} \, \vec{\beta}' P_\beta  \vec{\beta}
\text{~~with~~} P_\beta = D'_\beta D_\beta + \epsilon I_{K^*},
\end{align*}
are added to the log-likelihood to define an estimation criterion for
the spline parameters. A small ridge penalty is added in the
definition of $P_\beta$ to include the identifiability constraint on
$\vec\beta$. In Bayesian terms, it translates into priors on the
spline coefficients:
\begin{align*}
& p(\vec\gamma|\kappa_\gamma) \propto \kappa_\gamma^{\rho(P_\gamma)/2}
\exp\left(-{\kappa_\gamma\over 2} \vec\gamma'P_\gamma\vec\gamma \right) ;
&  p(\vec\beta|\kappa_\beta) \propto \kappa_\beta^{K^{*}/2}
\exp\left(-{\kappa_\beta\over 2} \vec\beta'P_\beta\vec\beta \right).
\end{align*}
With gamma priors on the penalty coefficients,
$\kappa_\gamma\sim\G{a_\gamma}{b_\gamma}, \kappa_\beta\sim\G{a_\beta}{b_\beta}$,
one can show that the marginal posterior for $(\vec\gamma,\vec\beta)$ is
\begin{align}
& p(\vec\gamma,\vec\beta|\D) \propto 
{L(\vec\gamma,\vec\beta|\D) 
   \over 
\left(b_\gamma +{1\over 2}\,\vec\gamma'P_\gamma\vec\gamma
\right)^{a_\gamma+{\rho(P_\gamma) \over 2}}
\left(b_\beta +{1\over 2}\,\vec\beta'P_\beta\vec\beta \right)^{a_\beta+{\rho(P_\beta) \over 2}}}
\label{MarginalPosterior:Eq2}
\end{align}
MAP estimates $(\hat{\vec\gamma},\hat{\vec\beta})$ for the splines
coefficients can be obtained by maximizing the last expression. Again,
an importance sampler based on a normal approximation could be set up
to explore the joint posterior of the spline parameters. Given the
increase in the number of parameters, we prefer to use a
block-Metropolis algorithm where vectorial proposals are made
sequentially for $\vec\gamma$ and $\vec\beta$. Our experience shows
that the following algorithm is very efficient:
\begin{itemize}
\item[--] Obtain an estimation $(\hat{\vec\gamma},\hat{\vec\beta})$ of
  the posterior mode and an approximation
  $(\hat{\vec\sigma}_\gamma,\hat{\vec\sigma}_\beta)$ to their marginal
  standard errors from (the diagonal of minus the inverse of) the
  Hessian matrix at the mode. Let
  $\hat\Sigma_\gamma=\text{diag}(\hat{\vec\sigma}_\gamma^2)$,
  $\hat\Sigma_\beta=\text{diag}(\hat{\vec\sigma}_\beta^2)$.
\item[--] At iteration $m$,  
  \begin{enumerate}
    \item Generate a proposal $\tilde{\vec\gamma}$ from the
      multivariate normal distribution \\
   ${\cal N}_{K}\left(\vec\gamma^{(m-1)},\varsigma_\gamma\,
      \hat\Sigma_\gamma\right)$ and $u$ from a
    uniform distribution ${\cal U}_{(0,1)}$. Let
$\varrho=\min\left\{1,{p(\tilde{\vec\gamma},\vec\beta^{(m-1)}|\D)
 \over
 p(\vec\gamma^{(m-1)},\vec\beta^{(m-1)})}\right\}$.
Set ${\vec\gamma}^{(m)}=\tilde{\vec\gamma}$ if $u\leq \varrho$
and ${\vec\gamma}^{(m)}={\vec\gamma}^{(m-1)}$ otherwise.
    \item Generate a proposal $\tilde{\vec\beta}$ from the
      multivariate normal distribution \\
   ${\cal N}_{K^*}\left(\vec\beta^{(m-1)},\varsigma_\beta \,
      \hat\Sigma_\beta\right)$ and $u$ from a
    uniform distribution ${\cal U}_{(0,1)}$. Let
$\varrho=\min\left\{1,{p(\vec\gamma^{(m)},\tilde{\vec\beta}|\D)
 \over
 p(\vec\gamma^{(m)},\vec\beta^{(m-1)}|\D)}\right\}$.
Set ${\vec\beta}^{(m)}=\tilde{\vec\beta}$ if $u\leq \varrho$
and ${\vec\beta}^{(m)}={\vec\beta}^{(m-1)}$ otherwise.
  \end{enumerate}
\item[--] The algorithm starts with $(\vec\gamma^{(0)},\vec\beta^{(0)})
  =(\hat{\vec\gamma},\hat{\vec\beta})$. The values of
  $\varsigma_\gamma$ and $\varsigma_\beta$ are tuned automatically
  during the burn-in to have acceptance probabilities around $0.20$
  \citep{Haario:Adapt:01}. The matrices $\hat\Sigma_\gamma$ and
  $\hat\Sigma_\beta$ are updated half-way during the burnin using the
  empirical variance-covariance of the corresponding chains.
\end{itemize}
The generated chain
$\left\{(\vec\gamma^{(m)},\vec\beta^{(m)}):m=1,\ldots,M\right\}$ can
be used to estimate a (simultaneous) credible region for any function
of the conditional copula generator. For example, if one is interested
in such a region for the conditional Kendall's tau, compute
\begin{align}
  &\tau^{(m)}(x_j)=1+4\int_0^1 \lambda_{\gamma^{(m)},\beta^{(m)}}(s|x_j)
\,ds \label{ConditionalKendallTau:PowerAlt:Eq}
\end{align}
on a fine grid of values $\left\{x_j\right\}_{j=1}^J$ on $(0,1)$.  The
so-obtained $M$ trajectories
$$\left\{\left(x_j,\tau^{(m)}(x_j)\right):j=1,\ldots,J\right\}_{m=1}^M$$
can be used to estimate pointwise or simultaneous \citep{Held:04}
credible regions for $\tau(\cdot)$.

\subsubsection{Simulation study}
  \label{PowerAltCopula:SimulationStudy:Sect}  
We have chosen to report the results of a study
where the data are simulated from a conditional copula outside the
additive conditional spline Archimedean copula family to assess the
ability of that tool to model the dynamics of specific dependence
structures. 
More specifically, for given covariate values $\{x_i:i=1,\ldots,n\}$
uniformly distributed on $(0,1)$,
we have simulated a data pair $(u_i,v_i)$ from a Clayton or from a
Frank copula with a dependence parameter such that the corresponding
Kendall's tau is
$$\tau(x_i) = .5 + .3 \sin\left(1.6 \pi x_i^{1.5}\right),$$
yielding values for tau oscillating between .2 and .8, a particularly
challenging set-up. The considered samples sizes are $n=250$, $500$
and $2000$. $S=500$ replicates were considered.

\begin{figure}\centering
\begin{tabular}{cc}
\includegraphics[width=7cm]{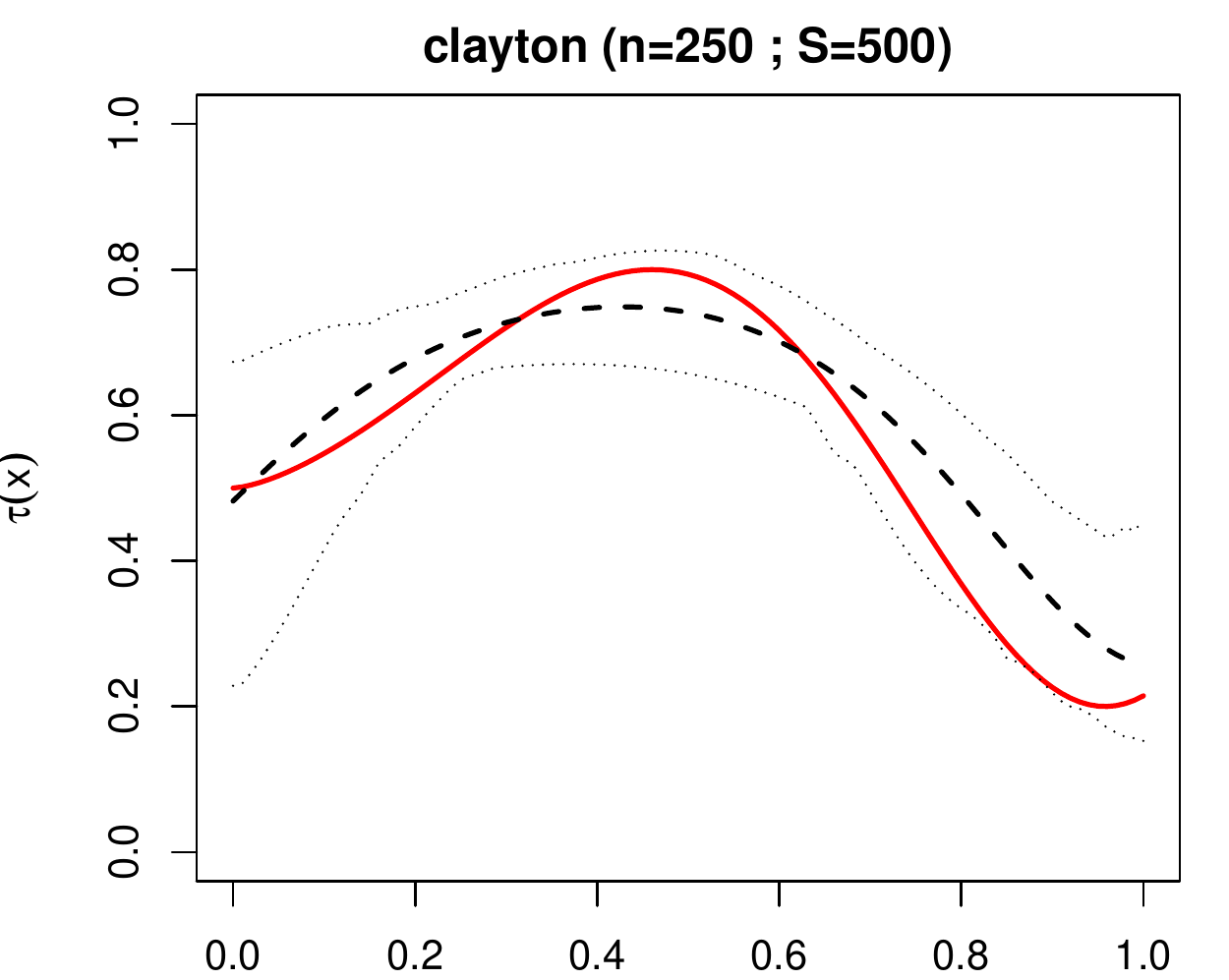} &
\includegraphics[width=7cm]{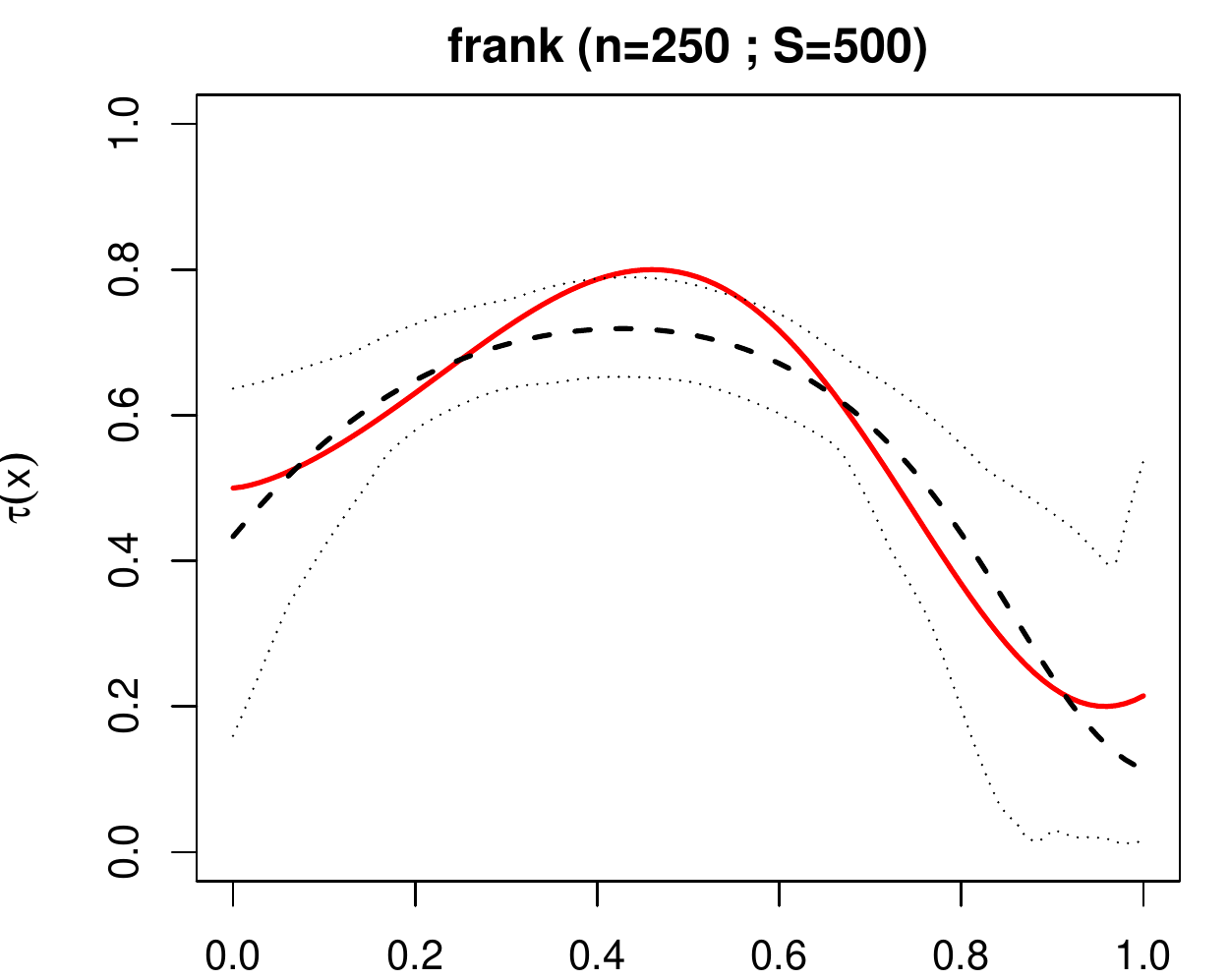} \\
\includegraphics[width=7cm]{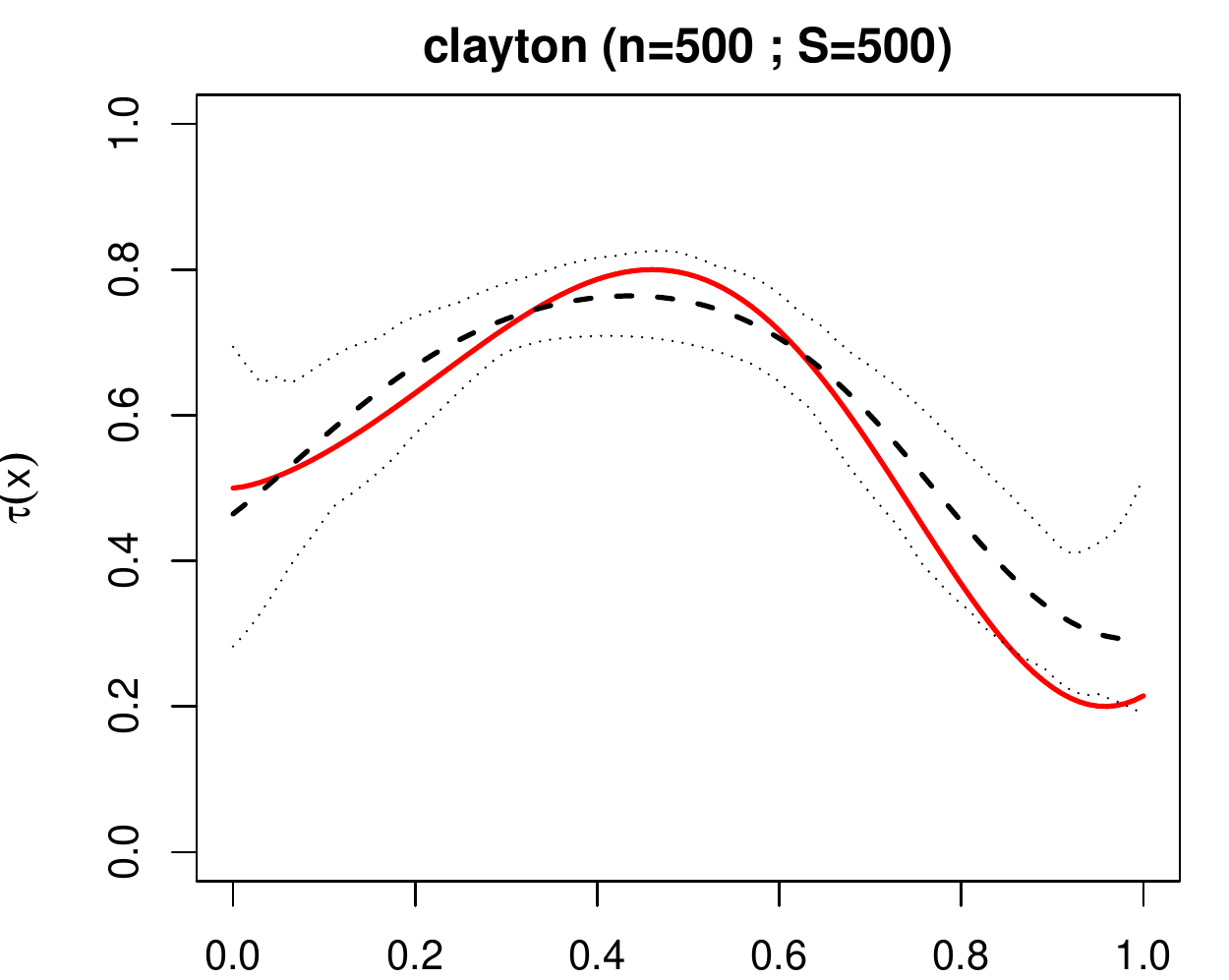} &
\includegraphics[width=7cm]{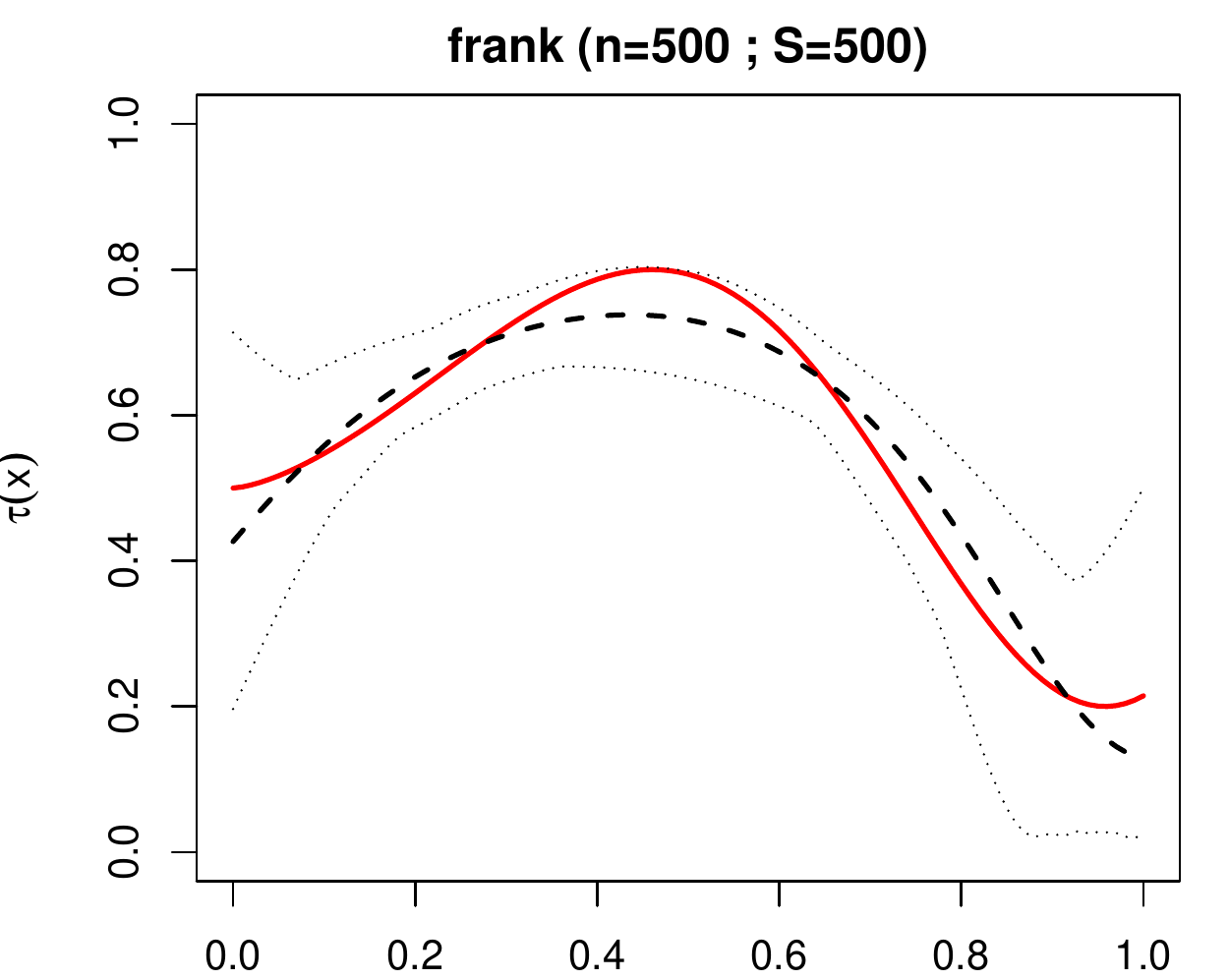} \\
\includegraphics[width=7cm]{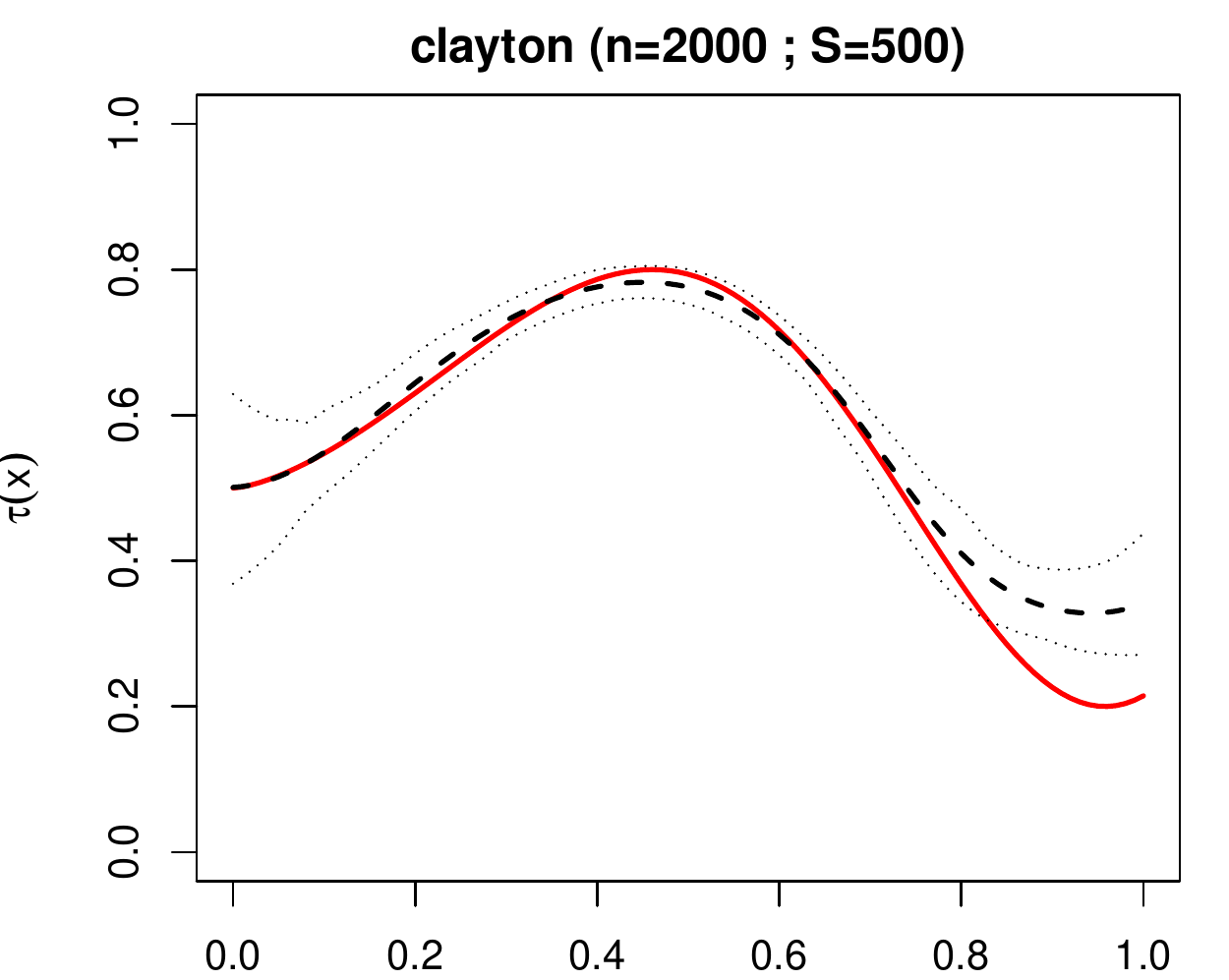} &
\includegraphics[width=7cm]{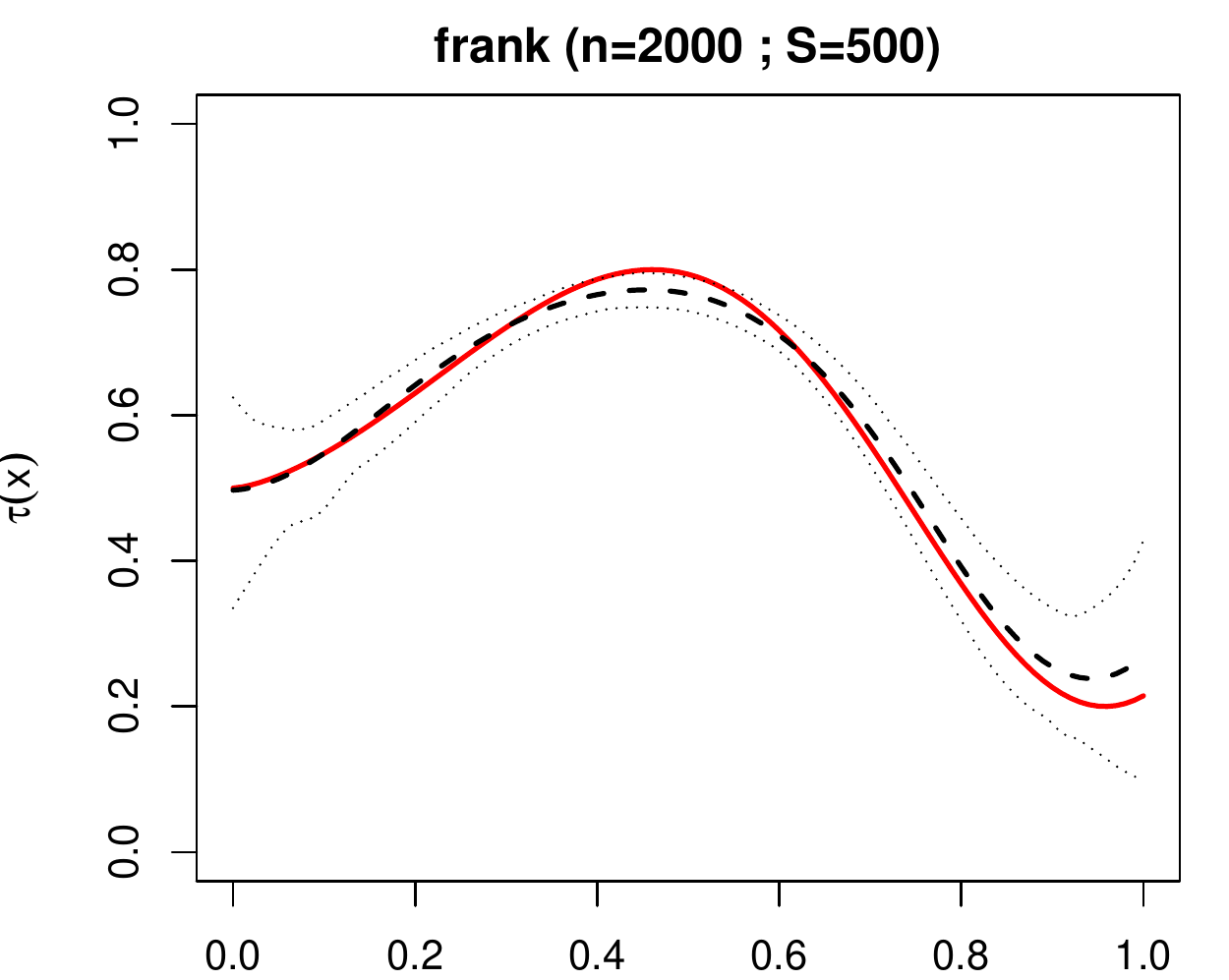}
\end{tabular}
\caption{\label{simulationPowerAlt:Fig} Estimate of the conditional
  Kendall's tau corresponding to the additive conditional spline
  Archimedean copula family fitted to $n$ data generated from a
  Clayton or a Frank copula with varying Kendall's tau: true Kendall's
  $\tau(x)$ (solid), mean (dashed line) of the $S$ estimated $\tau(x)$
  and envelope (dotted lines) containing 95\% of the $\tau(x)$
  estimates over the $S=500$ replicates.}
\end{figure}

Given that the data were simulated outside the assumed model family
and the demanding simulation set-up, we did not expect a perfect
reconstruction of the underlying dependence structure, but hoped that
reasonable estimations for the changing strength of dependence would
be obtained by forcing that model.  This is indeed the conclusion that
can be drawn by inspecting Fig.~\ref{simulationPowerAlt:Fig} where the
mean (over the $S$ replicates) of the estimated conditional Kendall's
tau (see (\ref{ConditionalKendallTau:PowerAlt:Eq})) corresponding
to the fitted additive conditional spline Archimedean copula model can
be compared to the function $\tau(x)$ used to simulate the
data. Whatever the underlying copula or sample size, the global
evolution of Kendall's tau with the covariate is correctly captured,
with results improving with sample size. The bias becomes negligible
for large $n$, except perhaps for the largest values of $x$ when the
data are generated using a Clayton copula with a small Kendall's tau.

\subsection{The flex-power Archimedean  copula family}
 \label{FlexPowerFamily:Sect}
\begin{figure}[bt!]\centering
 \includegraphics[width=13.5cm]{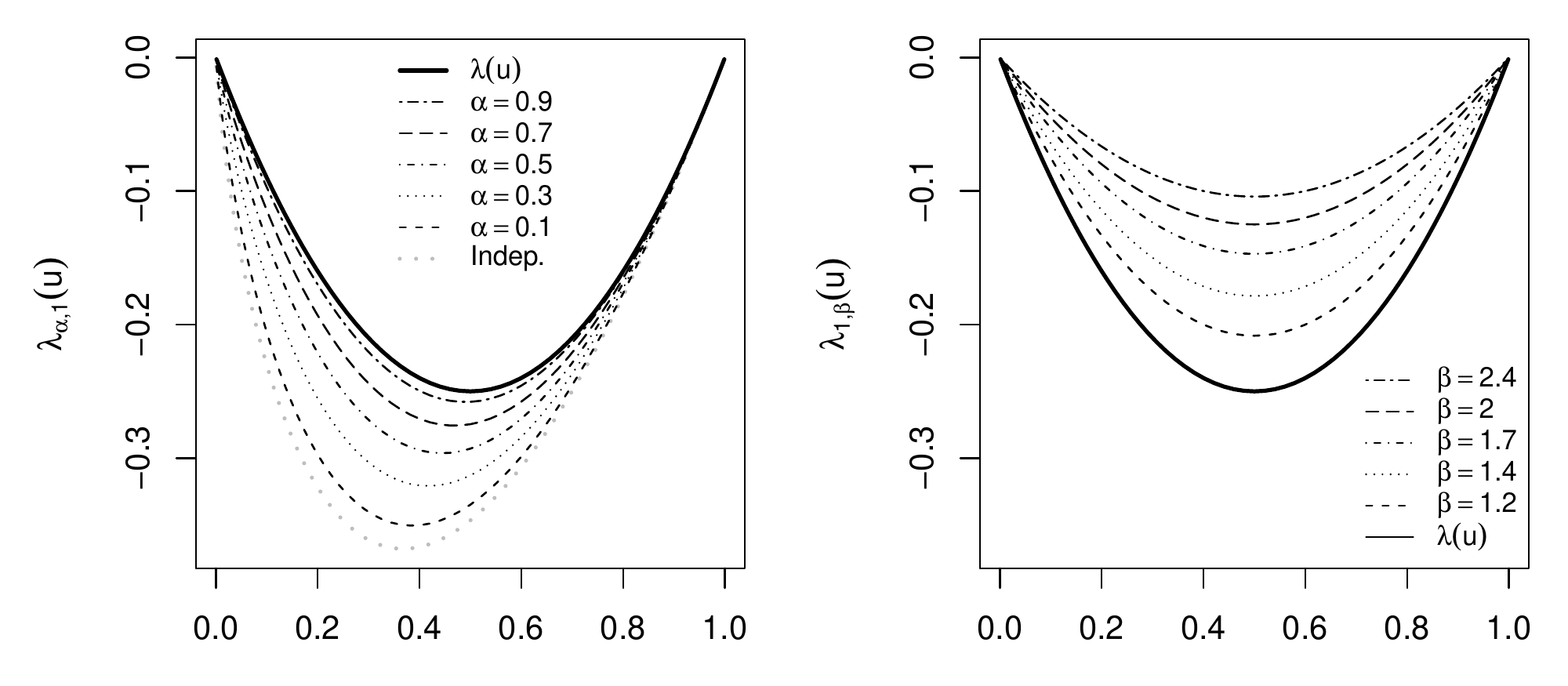}
 \caption{\label{lambert:LambdaPowerEffects}
   Effect of $\alpha$ and $\beta$ on the lambda function in the power
   Archimedean family (for a Clayton($\theta=1$) reference generator).}
\end{figure}

 Other approaches were investigated. One is based on (what we suggest
 to name) the {\em power Archimedean copula family}.  It relies on the
 following result \citep[see e.g.][]{NelsenR:Intr:99}:
 if $\varphi(\cdot)$ is an Archimedean copula generator, then
\begin{enumerate}
\item {\em Interior power transform}:\\
  $\varphi_{\alpha,1}(t)=\varphi(t^{\alpha})$ is also a generator if
  $\alpha \in (0,1]$ ;
\item {\em Exterior power transform}:\\
  $\varphi_{1,\beta}(t)=\left(\varphi(t)\right)^{\beta}$ is also a
  generator if $\beta \geq 1$.
\end{enumerate}
The effect of $\alpha$ and $\beta$ on a reference generator
$\varphi(t)$ (with Kendall's tau $\tau_0$) is illustrated on
Fig.~\ref{lambert:LambdaPowerEffects} when starting from a
Clayton($\theta=1$) generator.
Remembering the relationship between Kendall's tau and the area
decribed by the lambda function and the horizontal axis, see
(\ref{KendallTauAndLambda:Eqn}), one can see 
that Kendall's tau increases with $\alpha$ (with a maximal value
$\tau_0$ when $\alpha=1$, corresponding to the reference generator)
and with $\beta$ (with a minimal value $\tau_0$ when
$\beta=1$).

We suggest to consider both transforms to model changes of the copula
generator with covariates, i.e. to take
\begin{eqnarray}
\varphi_{\alpha,\beta}(t|x)=\left[\varphi(t^{\alpha(x)})\right]^{\beta(x)} 
\label{ConditionalGenerator:Eq}
\end{eqnarray}
with flexible forms for
\begin{itemize}
\item the reference generator $\varphi(\cdot)$, see
  (\ref{lambert:NonparametricGenerator})-(\ref{derivativeOfG:Eq}) ;
\item the interior power 
\begin{eqnarray}
\alpha=\alpha(x)=
\left[1+\left(\sum_{k=1}^{K^{*}}b_{k}^{*}(x)\alpha_{k}\right)^{2}\right]^{-1}~;
\label{AlphaSpline:Eq}
\end{eqnarray}
\item the exterior power\begin{eqnarray}
\beta=\beta(x)= 1+\left(\sum_{k}^{K^{*}}b_{k}^{*}(x)\beta_{k}\right)^{2},
\label{BetaSpline:Eq}
\end{eqnarray}
\end{itemize}
where $\{b_{k}^{*}(\cdot):k=1,\ldots,K^*\}$ denotes a cubic B-spline
basis on the covariate space (relocated and rescaled to take values in
$(0,1)$). That model includes $(K+2K^*)$ parameters.  Like for the
reference generator, penalties can be introduced to favor smooth
changes with $x$ for $\alpha(x)$ and $\beta(x)$, and, hence, for the
conditional copula generator $\varphi_{\alpha,\beta}(t|x)$. If
$\kappa_\alpha$ (resp.\,$\kappa_\beta$) denote the penalty
coefficients for the spline parameters $\vec\alpha$
(resp.\,$\vec\beta$) in $\alpha(x)$ (resp.\,$\beta(x)$) and $P_\alpha$
(resp.\,$P_\beta$) the corresponding penalty matrix, then one can
show, using gamma priors $\kappa_\alpha\sim{\cal
  G}(a_\alpha,b_\alpha)$, $\kappa_\beta\sim{\cal G}(a_\beta,b_\beta)$
and the same reasoning as in Section \ref{CopulaSpline:Sect}, that the
marginal posterior for the spline coefficients is
\begin{align}
p(\vec\theta,\vec\alpha,\vec\beta|\D) &
\propto \int_0^{+\infty} \int_0^{+\infty} \int_0^{+\infty} 
L(\vec\theta,\vec\alpha,\vec\beta|D) \, p(\vec\theta|\kappa)
\nonumber \\
&\hspace{10ex}
\, p(\vec\alpha|\kappa_\alpha) \, p(\vec\beta|\kappa_\beta)
\, p(\kappa) \, p(\kappa_\alpha) \, p(\kappa_\beta)\, d\kappa \, d\kappa_\alpha
\, d\kappa_\beta
\nonumber \\
&\propto {L(\vec\theta,\vec\alpha,\vec\beta|\D) \over
 \left(b +{1\over 2}\vec\theta'P\vec\theta \right)^{a+{\rho(P) \over 2}}
 \left(b_\alpha +{1\over 2}\vec\alpha'P_\alpha\vec\alpha \right)^{a_\alpha+{\rho(P_\alpha) \over 2}}
 \left(b_\beta +{1\over 2}\vec\beta'P_\beta\vec\beta \right)^{a_\beta+{\rho(P_\beta) \over 2}}
}
\label{MarginalPosteriorFlexPower:Eq}
\end{align}
MAP estimates $(\hat{\vec\theta},\hat{\vec\alpha},\hat{\vec\beta})$
for the splines coefficients can be obtained by maximizing the last
expression. An adaptive block-Metropolis algorithm can be set up by
mimicking the approach in Section \ref{PowerAltFamily:Inference:Sec}
to sample the joint posterior of the spline parameters.

\subsubsection{Simulation study}
 \label{FlexPowerCopula:SimulationStudy:Sect} 
When the data are generated from the power Archimedean family where
the interior and exterior powers are functions of a covariate $x$, one
can show, using simulations (not reported to save space) that the
flex-power family is able to estimate the underlying conditional
copula in a precise and nearly unbiased way for large sample
sizes. This is not surprising given the results in Section
\ref{CopulaSpline:SimulationStudy:Sect}. 

\begin{figure}\centering
\begin{tabular}{cc}
\includegraphics[width=7cm]{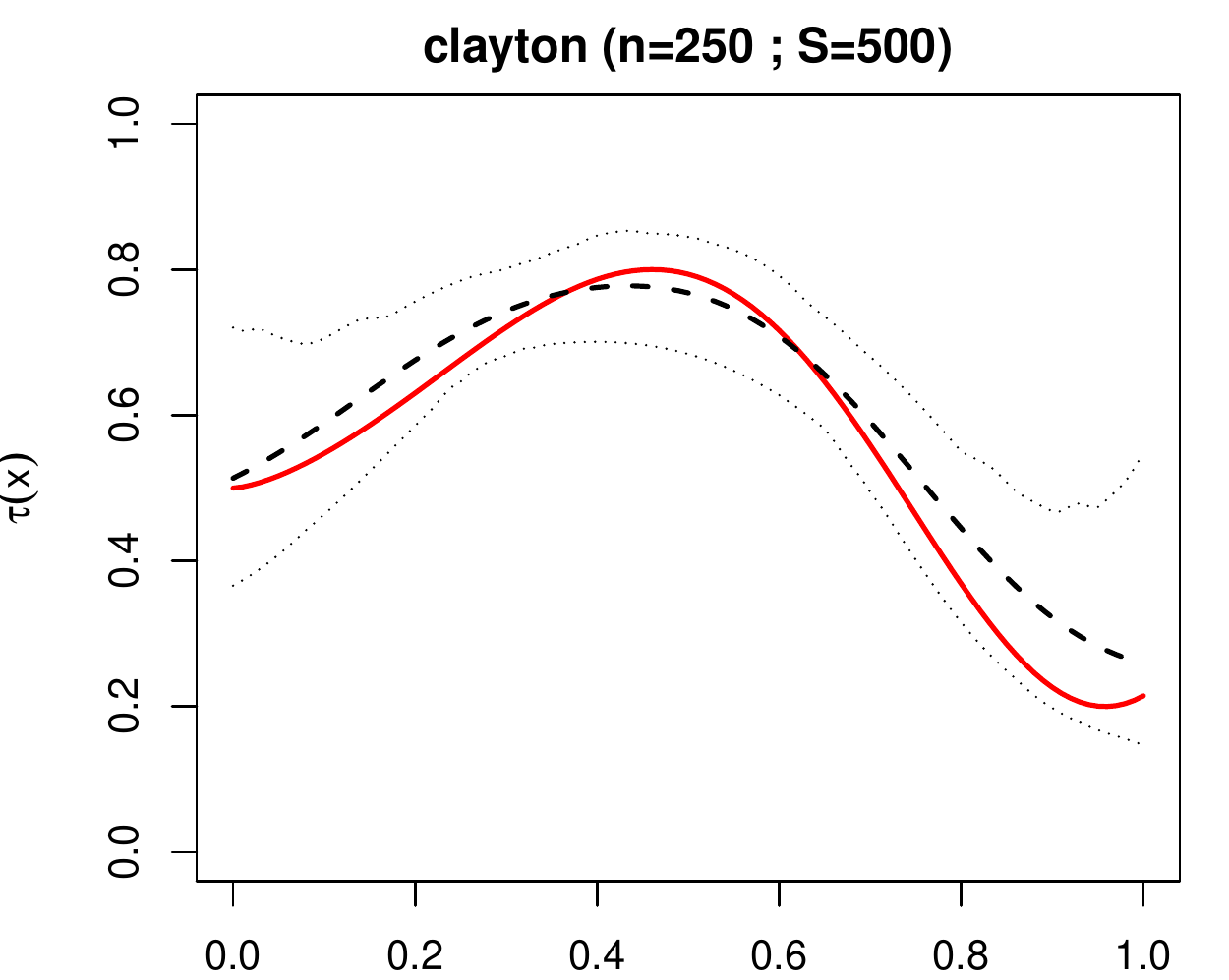} &
\includegraphics[width=7cm]{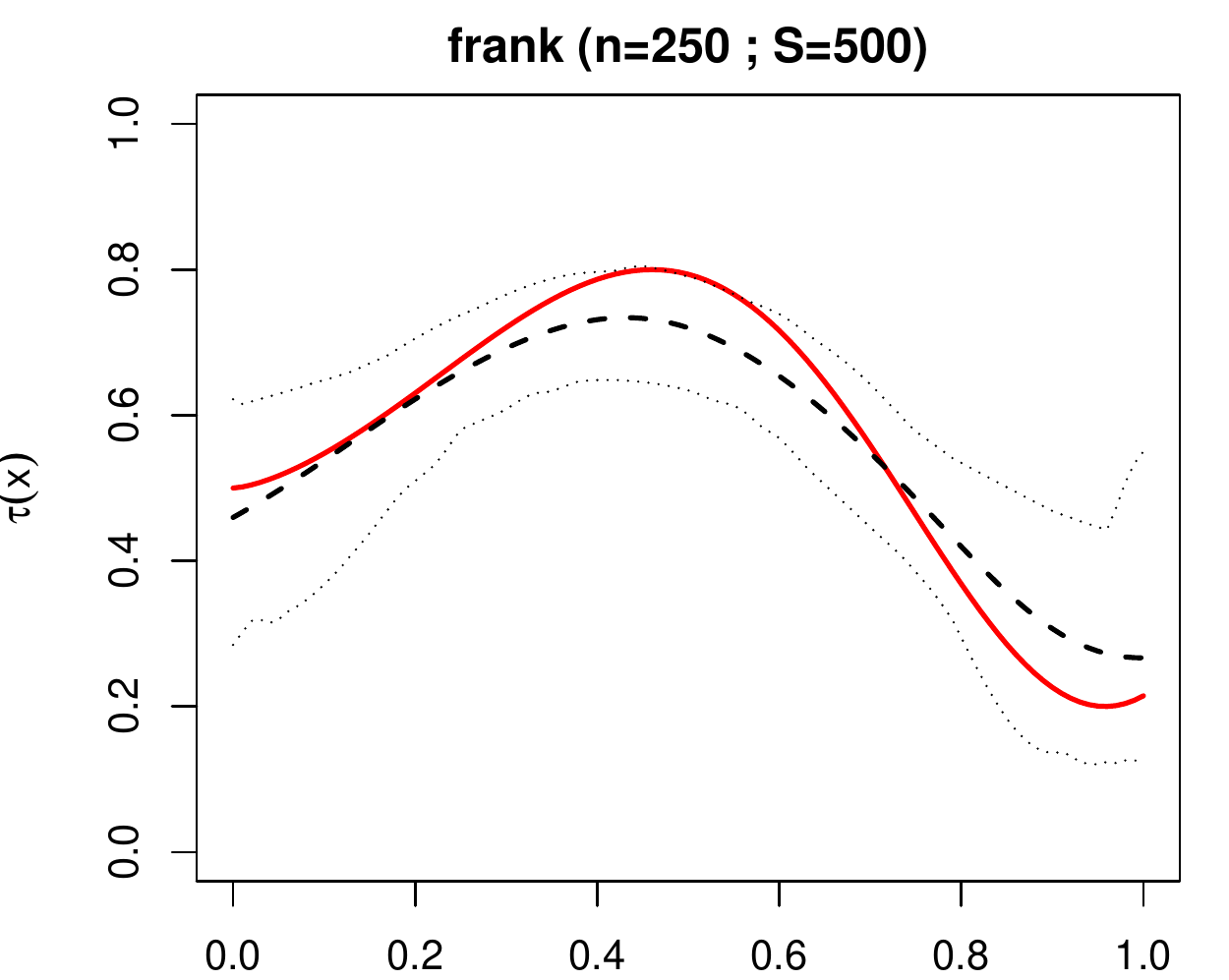} \\
\includegraphics[width=7cm]{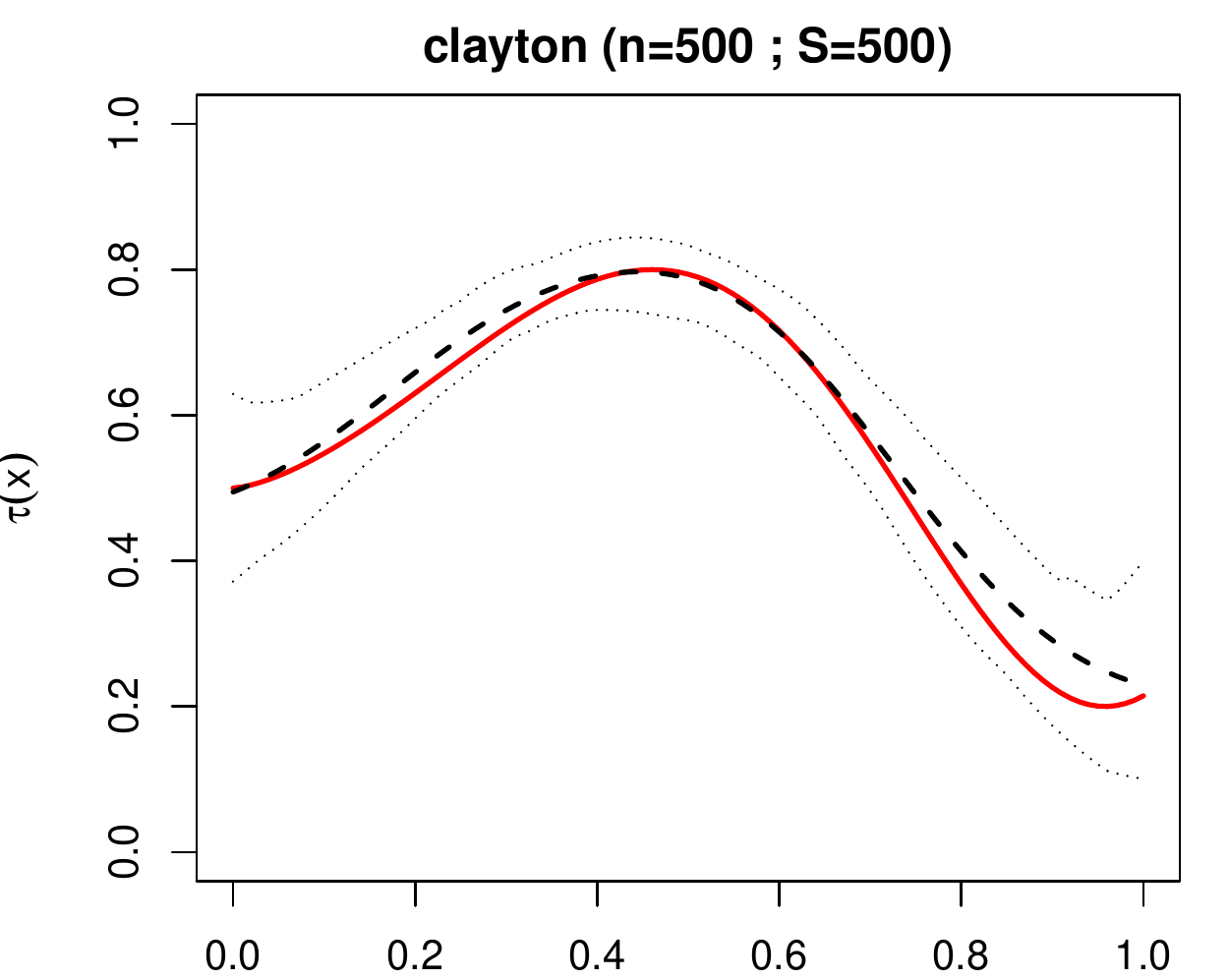} &
\includegraphics[width=7cm]{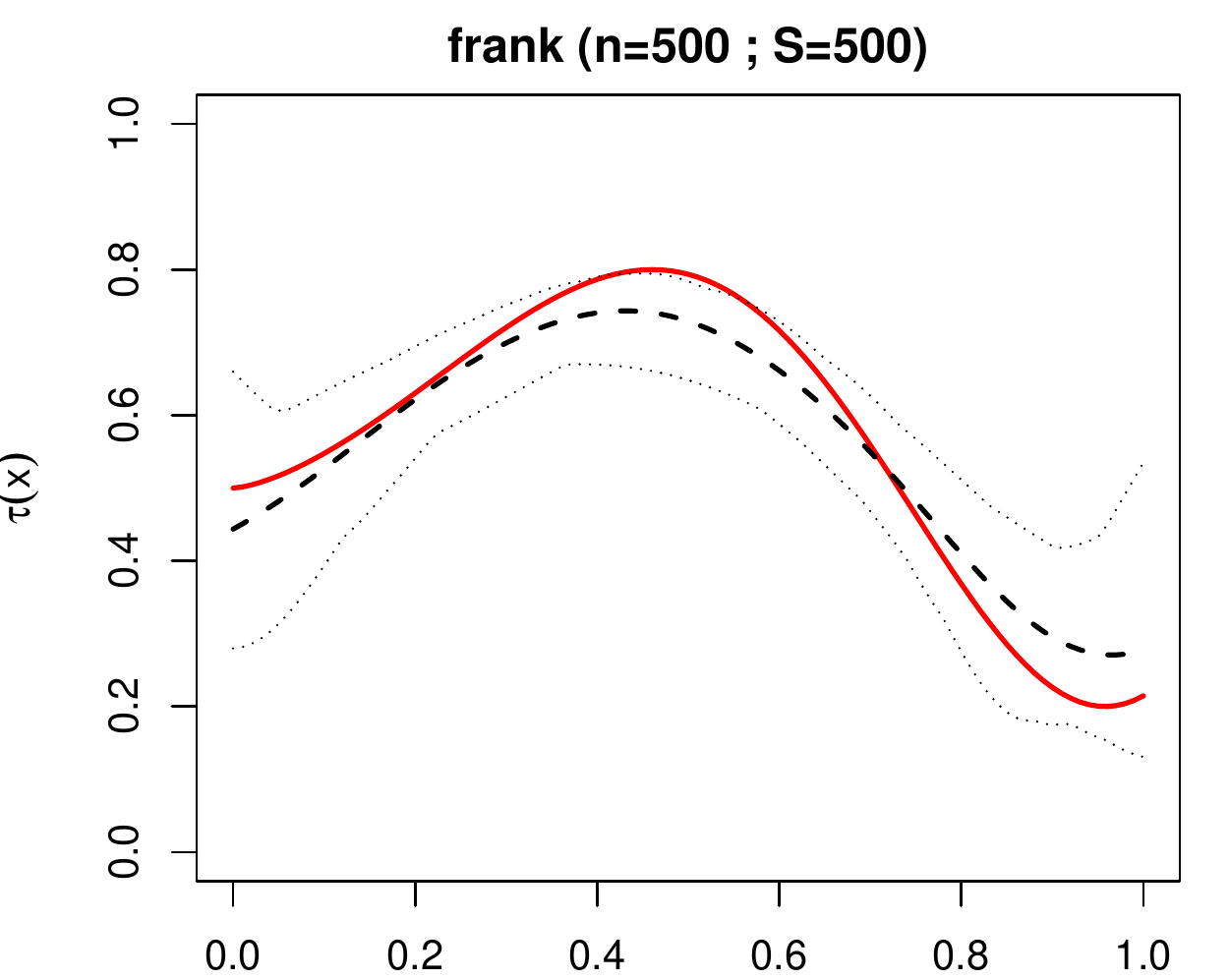} \\
\includegraphics[width=7cm]{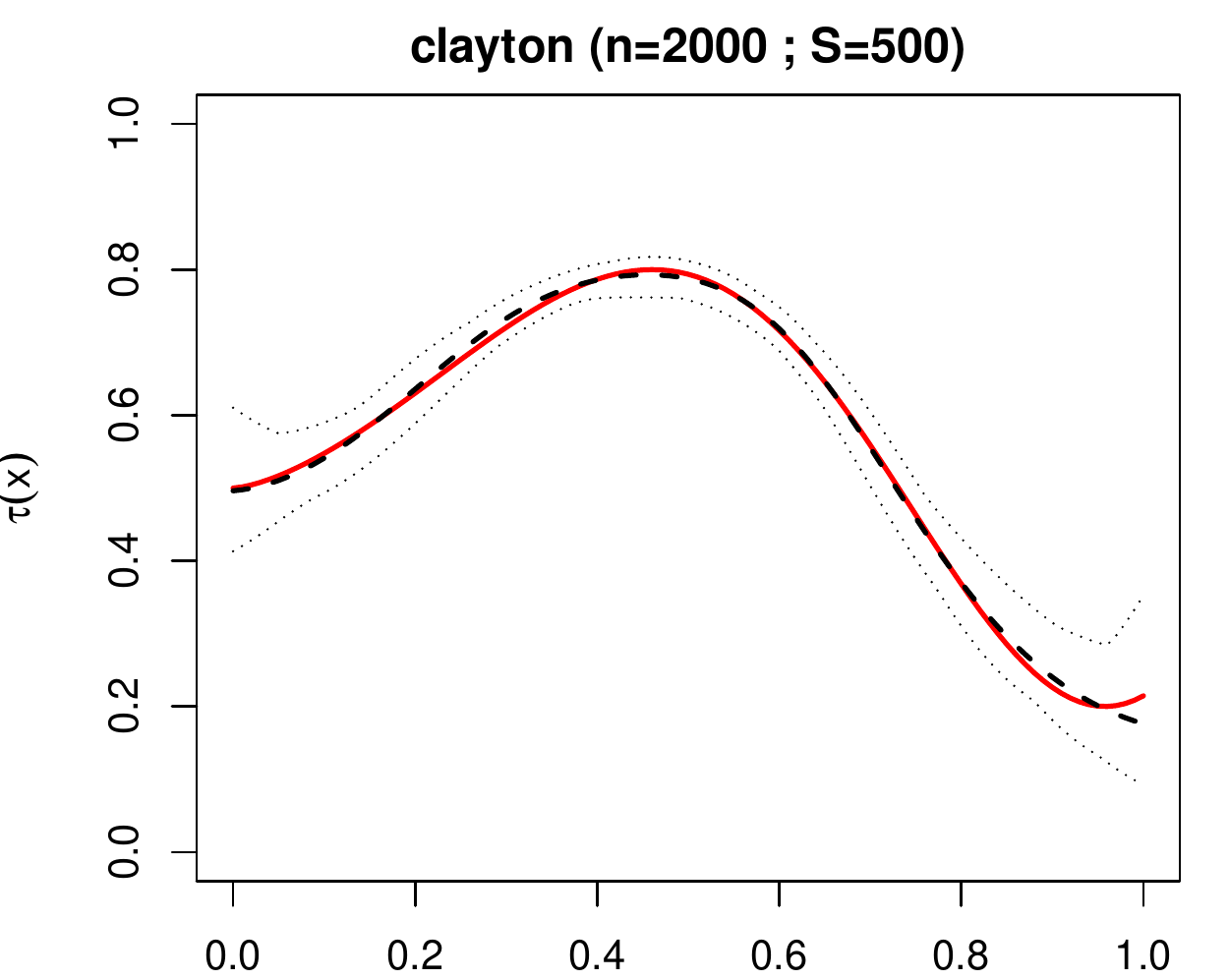} &
\includegraphics[width=7cm]{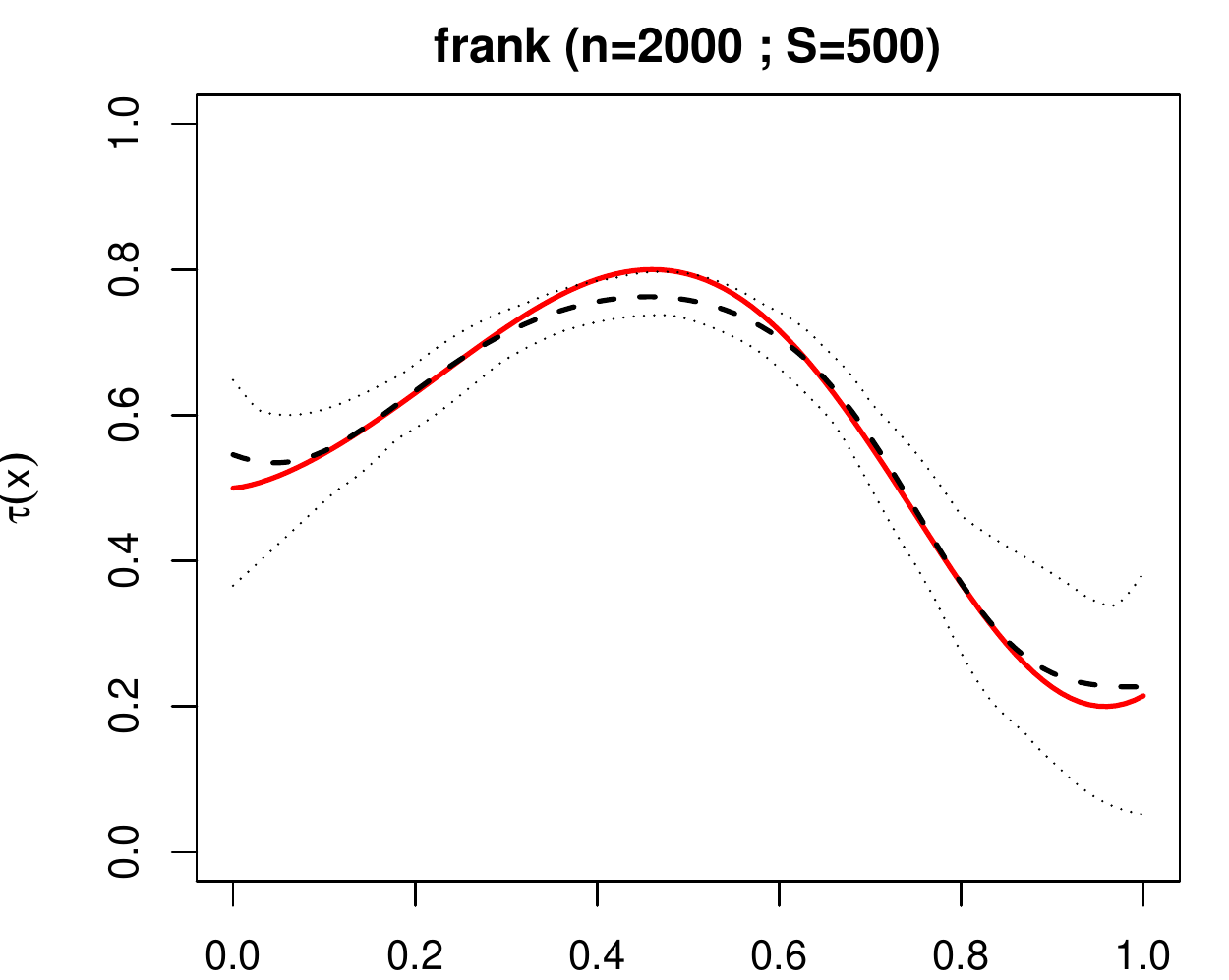}
\end{tabular}
\caption{\label{simulationPowerFamily:Fig} Estimate of the conditional
  Kendall's tau corresponding to a flex-power Archimedean copula
  fitted to $n$ data generated from a Clayton or a Frank copula with
  varying Kendall's tau: true Kendall's $\tau(x)$ (solid), mean
  (dashed line) of the $S$ estimated $\tau(x)$ and envelope (dotted
  lines) containing 95\% of the $\tau(x)$ estimates over the $S=500$
  replicates.}
\end{figure}

Instead, we report the results obtained when fitting the flex-power
Archime\-dean family to the same datasets as in Section
\ref{PowerAltCopula:SimulationStudy:Sect}.  Given that the data were
simulated outside the flex-power family, we did not expect a perfect
reconstruction of the underlying dependence structure, but like for
the additive model, hoped that reasonable estimations for the changing
strength of dependence would be obtained by forcing that model.  This
is indeed the conclusion that can be drawn by inspecting
Fig.~\ref{simulationPowerFamily:Fig} where the mean (over the $S$
replicates) of the estimated conditional Kendall's tau corresponding
to the fitted flex-power Archimedean copula can be compared to the
function $\tau(x)$ used to simulate the data. Whatever the underlying
copula or sample size, the bias for a given $x$ is small and decreases
with $n$. A comparison of Figs.\,\ref{simulationPowerAlt:Fig} and
\ref{simulationPowerFamily:Fig} suggests that, in the framework of the
simulation study, the flex-power family is a bit more performant than
the additive one, at the cost of $K^*$ extra (spline) parameter (per
covariate).

\section{Applications} \label{Application:Sec}
\subsection{Growth curves}
\begin{figure}\centering
\begin{tabular}{cc}
\includegraphics[width=6.5cm]{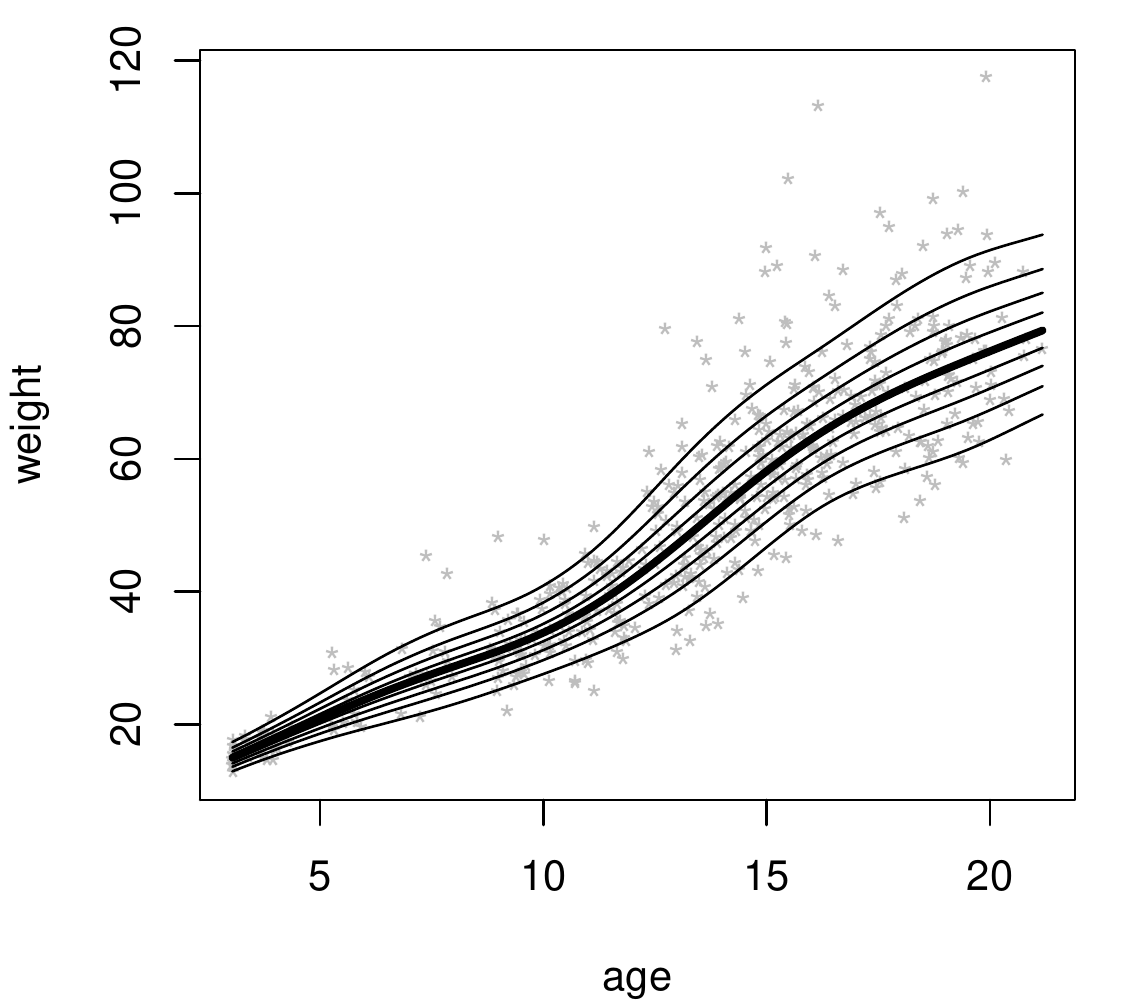} &
\includegraphics[width=6.5cm]{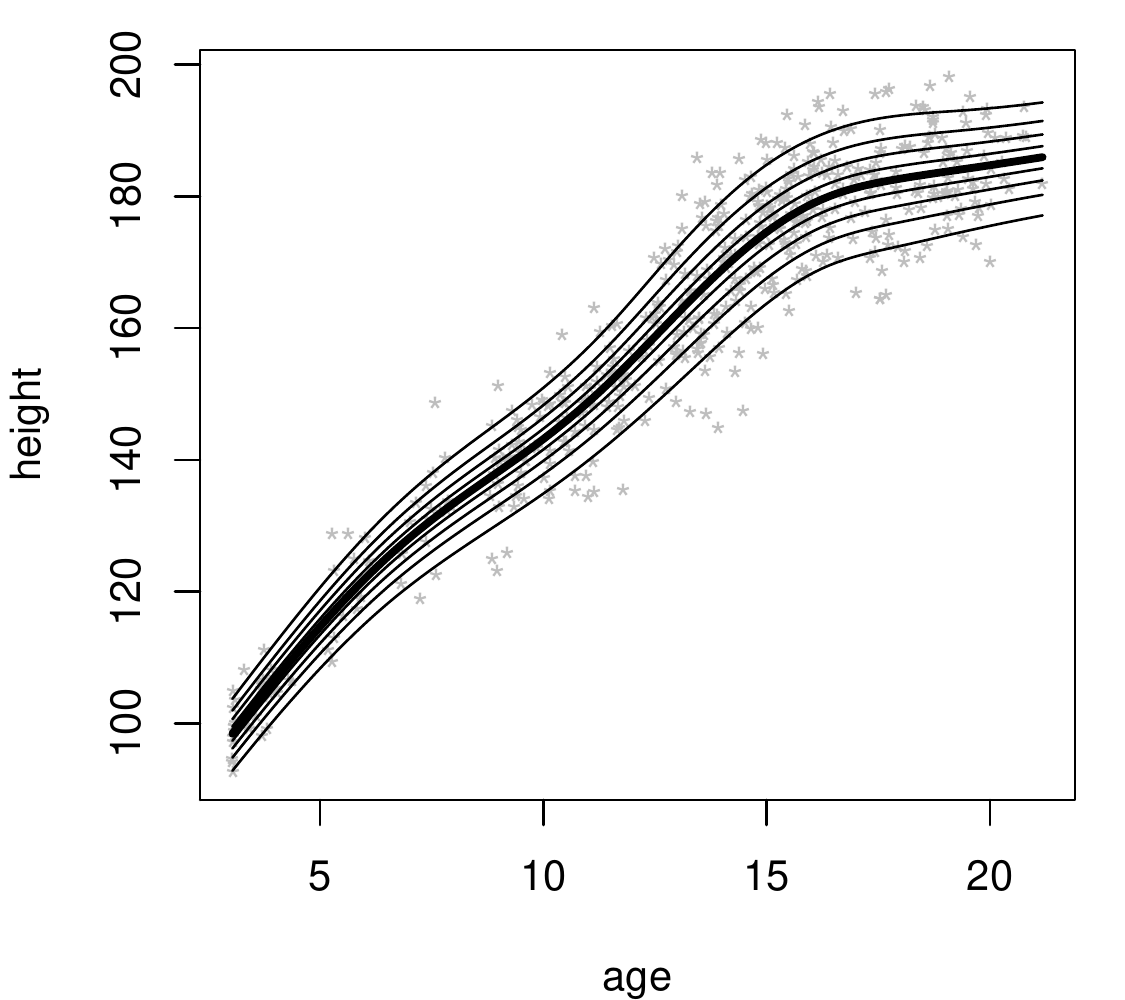}
\end{tabular}
\caption{\label{DutchBoys:FittedMarginalDeciles:Fig} Growth dataset:
  fitted conditional deciles for \ttf{Weight} and \ttf{Height} given
  \ttf{Age} using the flexible location-scale model in \citet{Lambert:2013}.}
\end{figure}

The additive conditional spline (Model 1) and flex-power (Model 2)
Archimedean copula models were also applied on the growth data
mentioned in Section 1 (see also
Fig.~\ref{lambert:DutchBoysCoplot}). It provides the height ($Y_1$)
and weight ($Y_2$) of $n=441$ young Dutch boys aged between 3 and
21. The margins were first modelled using the nonparametric additive
location-scale model described in \citet{Lambert:2013},
$$Y_j = \mu_j(x)+\sigma_j(x)\epsilon_j,$$
where $\mu_j(x)$ and $\sigma_j(x)$ denotes the conditional median and
inter-quartile range of $Y_j$ given \ttf{Age} ($=X$). The smooth
evolution of these quantities, as well as the pivotal distributions of
the $\epsilon_j$'s were described using penalized B-splines. The
resulting fitted conditional deciles for $Y_1$ and $Y_2$ are displayed
on Fig.\,\ref{DutchBoys:FittedMarginalDeciles:Fig}.  It reveals the
nonlinear relationship between the responses and \ttf{Age} as well as
their increasing dispersion with the latter variate.

\begin{table}
\centering
\begin{tabular}{lccc}
Copula model & \# par. & e.d. & DIC \\
\hline
Model 0 (Unconditional) & 11 & 3.2 & -310.4 \\
Model 1 (additive) & 16 & 6.0 & -319.0 \\
Model 2 (flex-power) & 21 & 7.0 & -318.4 
\end{tabular}
\wcaption{Growth dataset: effective number of parameters (e.d.) and
  DIC in the fitted copula models.}{Growth:DIC:Tab}
\end{table}

\begin{figure}[bt!]\centering
\begin{tabular}{cc}
\multicolumn{2}{c}{Additive conditional spline model (Model 1)}\\
 \includegraphics[width=6.5cm]{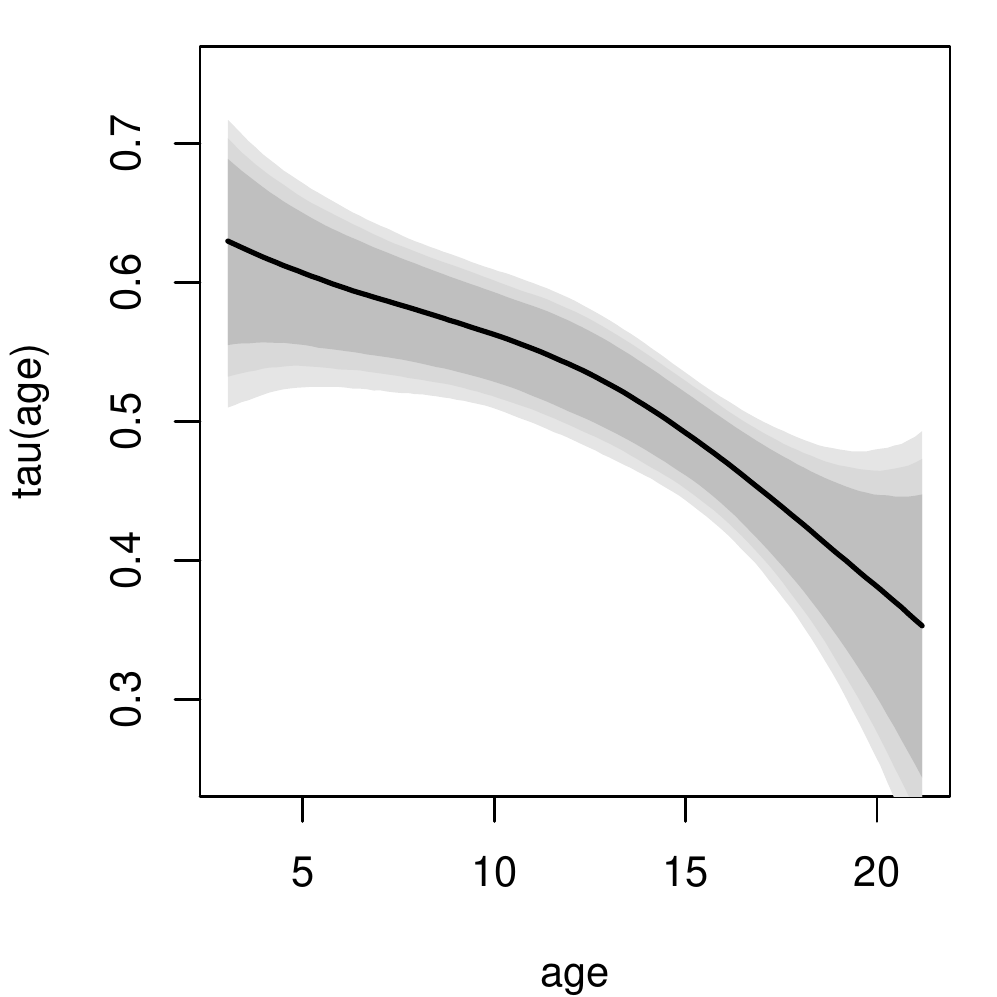} &
 \includegraphics[width=6.5cm]{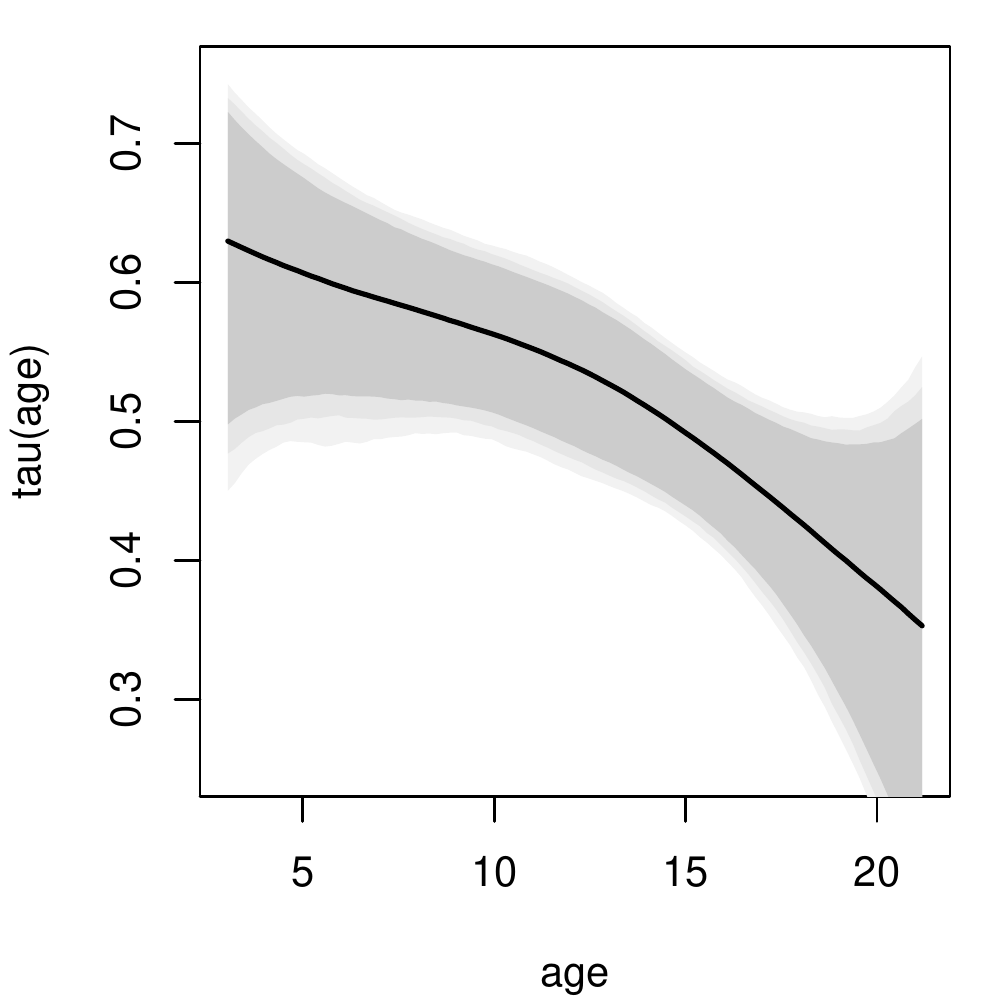} \\
\mbox{}\\
\multicolumn{2}{c}{Flex-power model (Model 2)}\\
 \includegraphics[width=6.5cm]{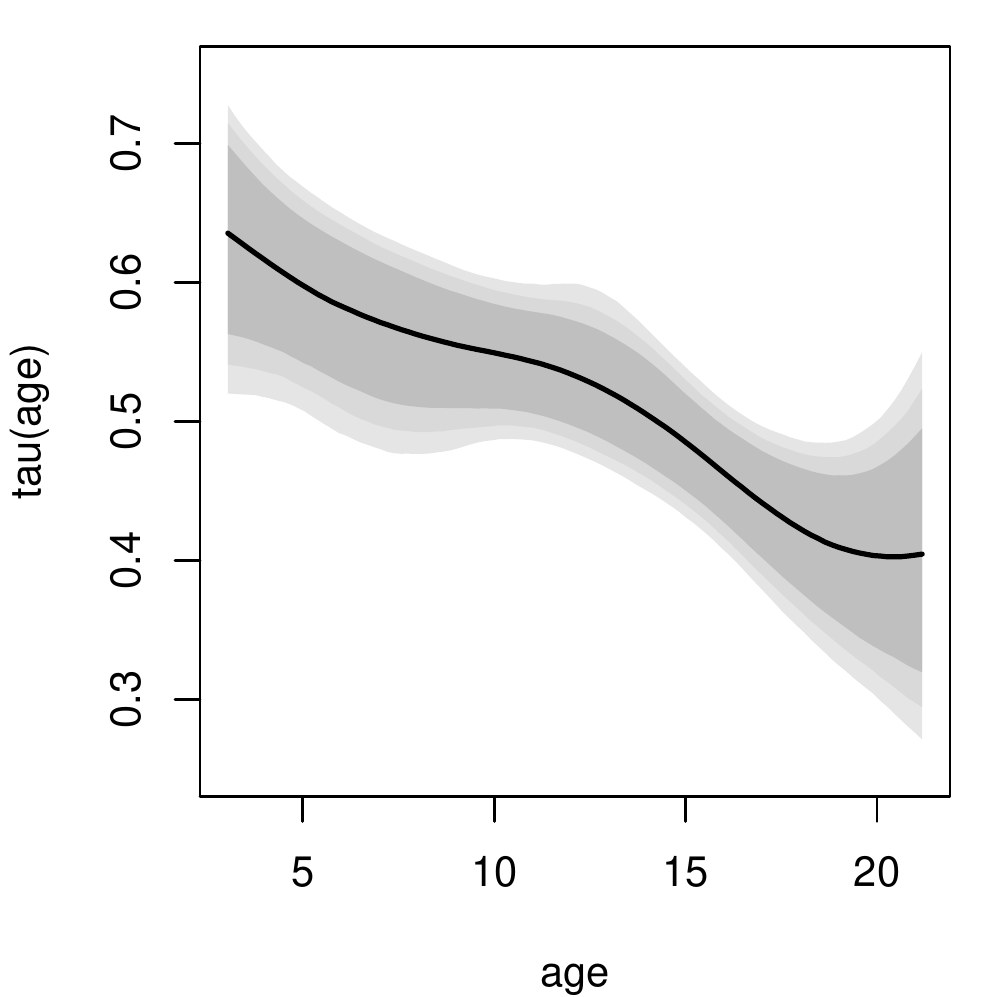} &
 \includegraphics[width=6.5cm]{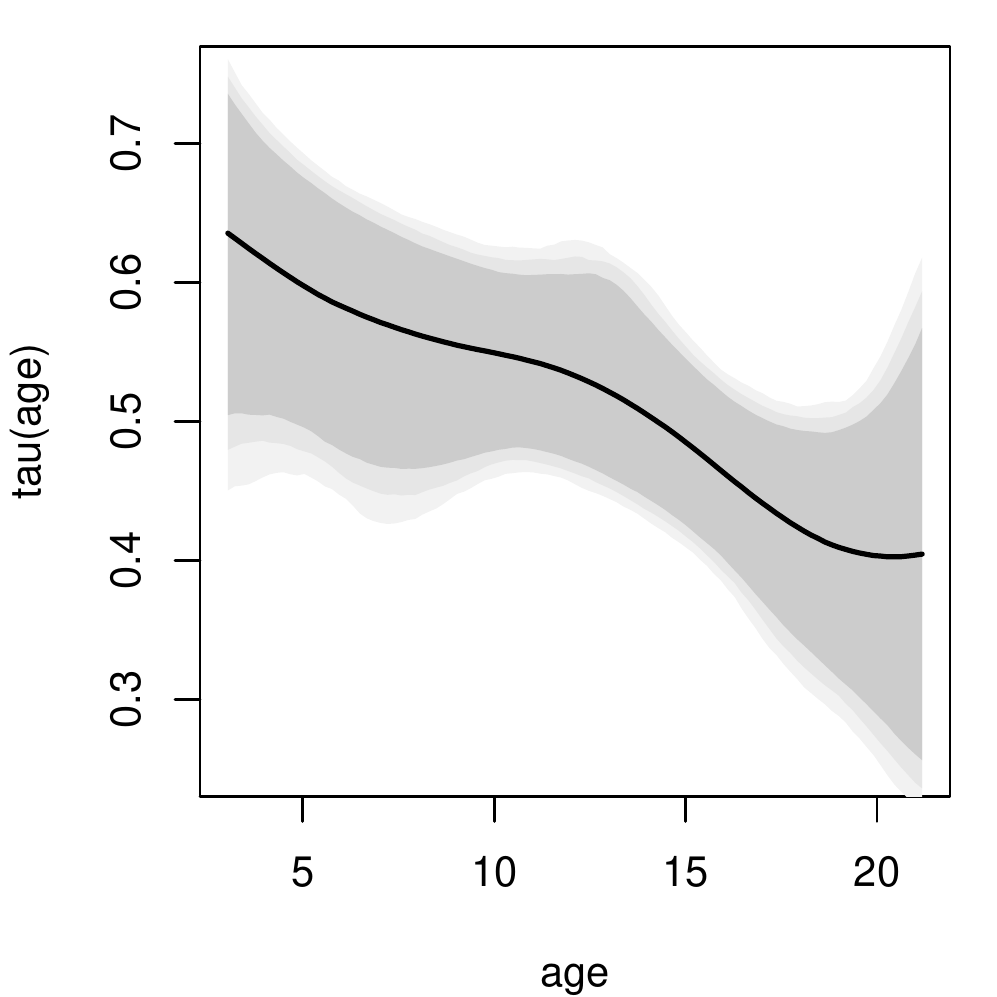}
\end{tabular}
 \caption{\label{DutchBoys:FittedTau:Fig} 
Growth dataset: fitted (solid line) conditional Kendall's tau under
the additive conditional spline and the flex-power Archimedean copula models. Grey areas corresponds
to .80, .90 and .95 pointwise (left panel) or simultaneous (right
panel) credible regions for $\tau$(\ttf{age}).}
\end{figure}

\begin{figure}[bt!]\centering
\includegraphics[width=12cm]{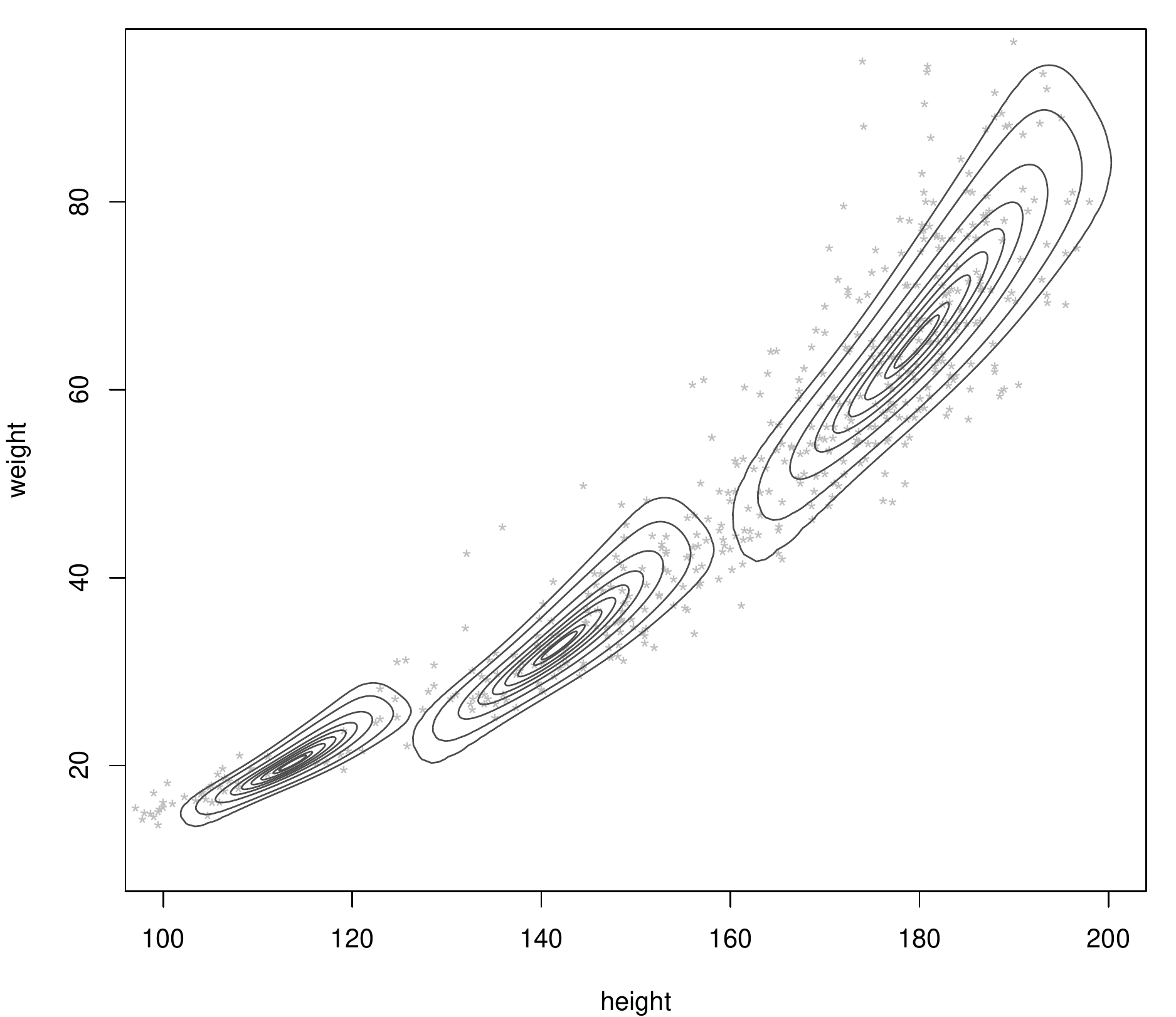}
\caption{\label{DutchBoys:FittedJoint:Fig} 
Growth dataset: contours of the fitted joint distribution for \ttf{Height} and
\ttf{Weight} at age 5, 10 and 17.}
\end{figure}

The unconditional and the two conditional copula models were adjusted
on the fitted conditional quantiles,
$$\left\{\left(x_i,\hat{u}_{1|x_i}=\hat{F}_1(y_{1i}|x_i),\hat{u}_{2|x_i}=\hat{F}_2(y_{2i}|x_i)\right):i=1,\ldots,n\right\},$$
using the methodology described in Sections \ref{CopulaSpline:Sect}
(Model 0), \ref{AdditiveCondSplineFamily:Sect} (Model 1) and
\ref{FlexPowerFamily:Sect} (Model 2), respectively, with $K=11$,
$K^*=5$ and $b=b_\alpha=b_\gamma=b_\beta=1$. Therefore, Model 0, Model
1 and Model 2 include $K=11$, $K+K^*=16$ and $K+2K^*=21$ parameters,
respectively (with identifiability constraints).  Adaptive
block-Metropolis algorithms with chains of length $30,000$ (and a
burnin of $1,000$) and initial values set at the MAP estimates were
used to explore the joint posterior of the spline parameters in the
two models. The deviance information criterions (DIC,
\citet{Spiegelhalter:2002wt}) are much smaller in Models 1 \& 2 than
in Model 0, suggesting a significant change of the strength of
association between the two responses with the covariate, see Table
\ref {Growth:DIC:Tab}. Smaller values were obtained for the DIC and
for the effective number of parameters in Model 1 (compared to Model
2). The posterior mean of the conditional Kendall's tau, also
estimated from the MCMC chains (see e.g.\,
(\ref{ConditionalKendallTau:PowerAlt:Eq}), is plotted with a 95\%
credible region in Fig.~\ref{DutchBoys:FittedTau:Fig}. It confirms and
quantifies the decreasing association between \ttf{weight} and
\ttf{height} suspected from Fig.~\ref{lambert:DutchBoysCoplot}. It is
very large at early childhood and decreases afterwards with an
apparent acceleration at the start of puberty.
The fitted joint distribution can also be visualized for any value of
the covariate, see Fig.\,\ref{DutchBoys:FittedJoint:Fig}, where
its contours for three values of \ttf{Age} are superposed to the
scatterplot of the whole dataset.

\subsection{Blood pressures and cholesterol level}
\begin{figure}[bt!]\centering
 \includegraphics[width=13cm]{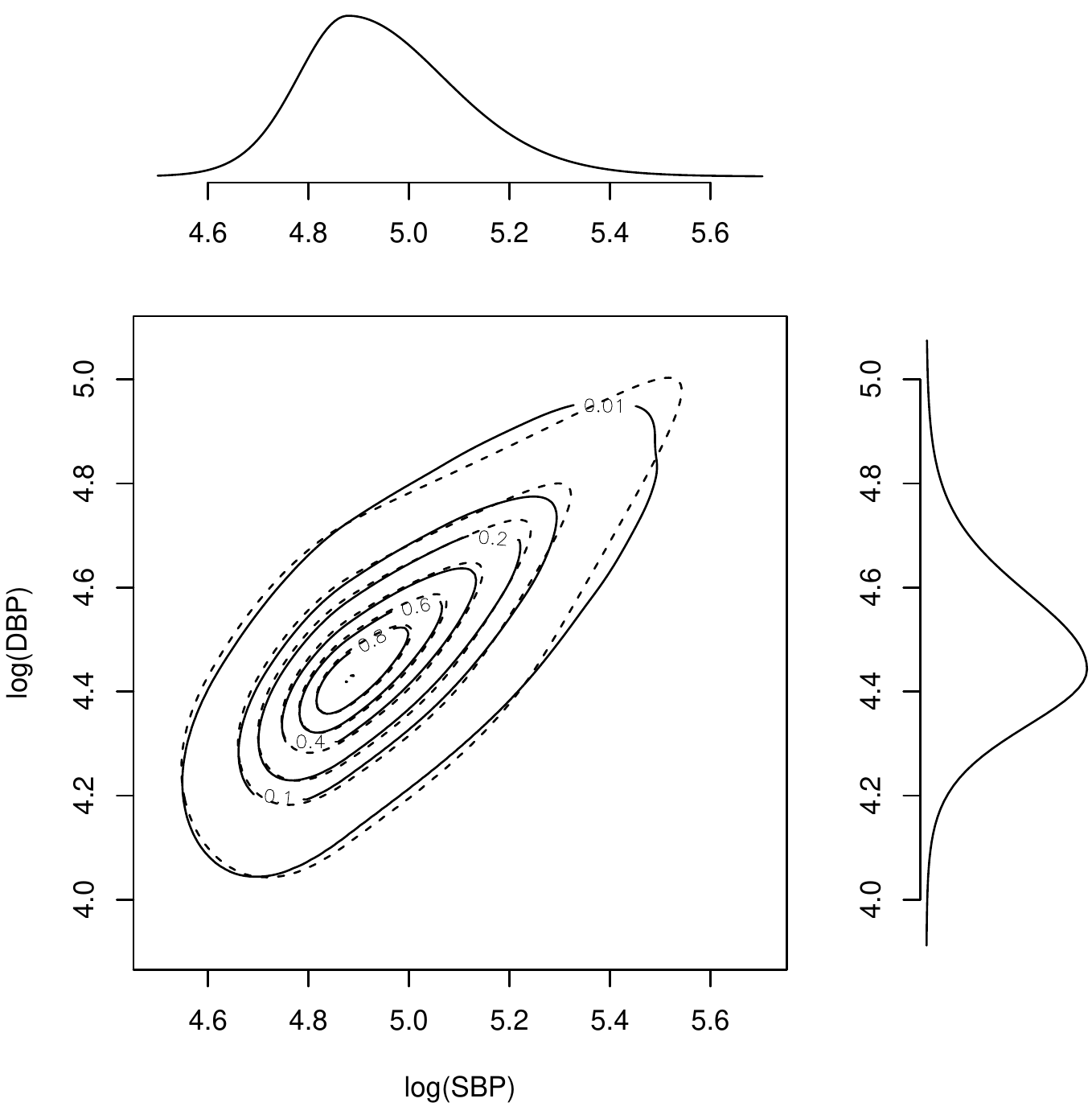}
 \caption{\label{FraminghamFig1}
	Fitted joint distribution for the log-diastolic and
        log-systolic blood pressures for an average cholesterol level
        using the flexible spline form (solid contours) or the Gumbel
        generator (dashed contours).}
\end{figure}

\begin{figure}[bt!]\centering
\begin{tabular}{cc}
{Additive conditional spline (Model 1)} & 
{Flex-power model (Model 2)} \\
 \includegraphics[width=6.5cm]{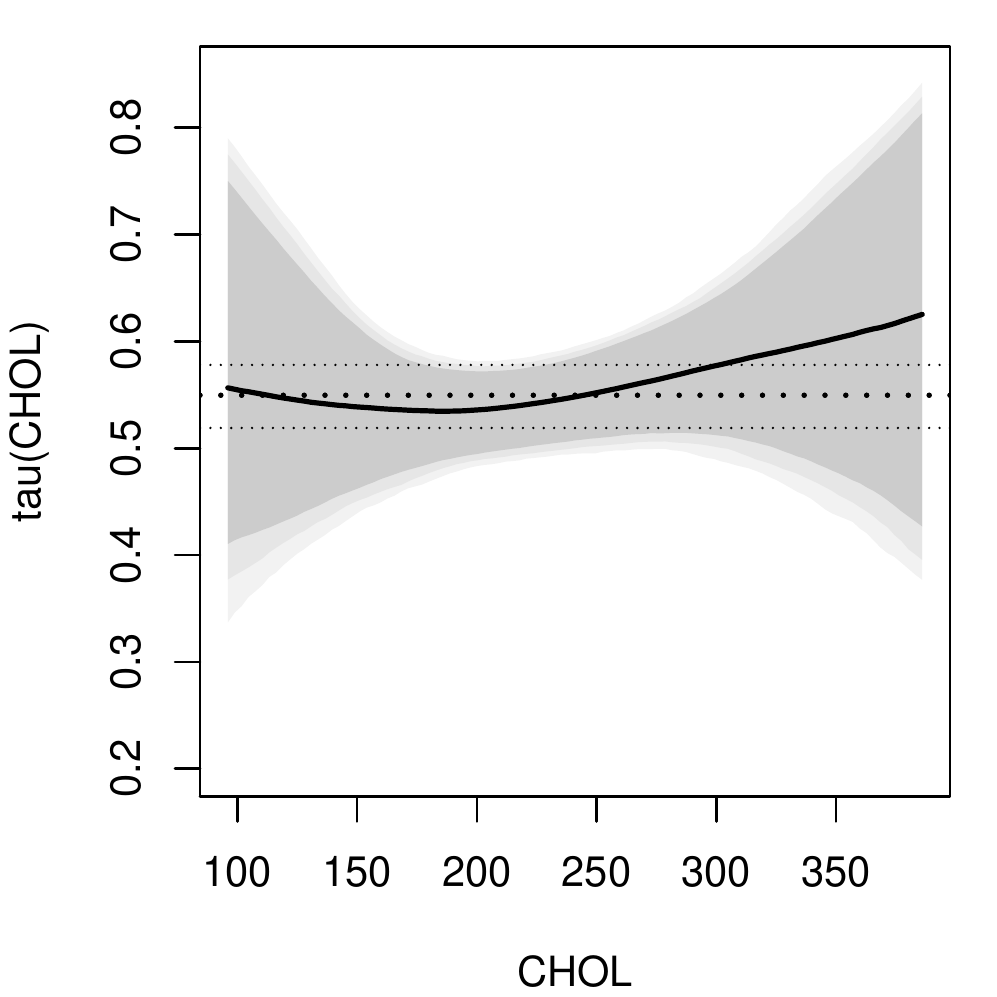} &
 \includegraphics[width=6.5cm]{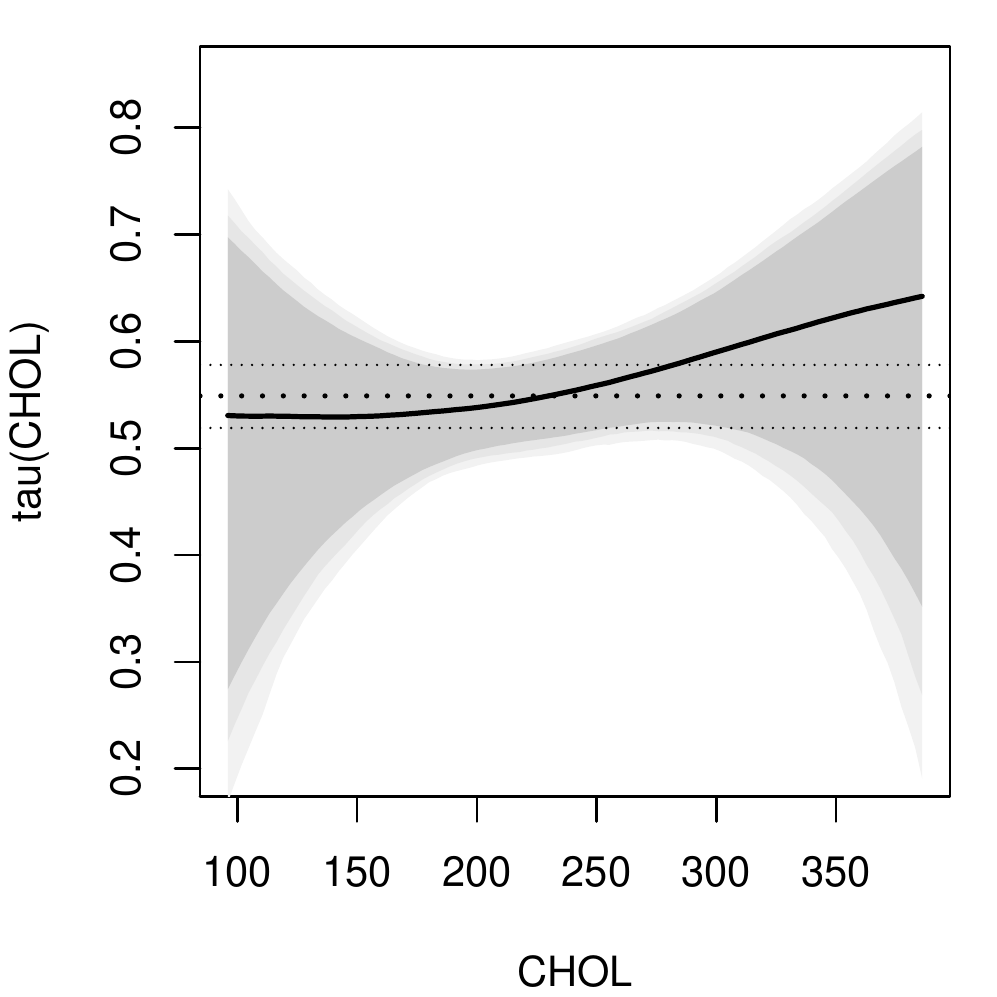} 
\end{tabular}
\caption{\label{BloodPressures:FittedTau:Fig} Blood pressure dataset:
  fitted (solid line) conditional Kendall's tau under the additive
  conditional spline and the flex-power Archimedean copula
  models. Grey areas corresponds to .80, .90 and .95 simultaneous
  credible regions for $\tau$(\ttf{CHOL}). The point estimate (dashed
  line) and the HPD region (dotted line) for Kendall's tau in the
  unconditional copula model are also provided.}
\end{figure}

The second example is based on data coming from the Framingham Heart
study data (\ttf{http://www.framingham.com/heart/}). We restrict our
attention to the evolution with the cholesterol level
($X=\text{CHOL}$) of the association between the diastolic
($Y_1=\text{DBP}$) and the systolic ($Y_2=\text{SBP}$) blood pressures
(in $mmHg$) measured on 663 male subjects at their first visit).  The
relation between each of the blood pressures with cholesterol was
quantified using regression models for the location of a 4-parameter
skewed Student distribution \citep{FernandezSteel:1998}, see Section 6
in \cite{Lambert:2007vo} for more details, yielding fitted marginal
quantiles
$$\left\{\left(x_i,\hat{u}_{1|x_i}=\hat{F}_1(y_{1i}|x_i),\hat{u}_{2|x_i}=\hat{F}_2(y_{2i}|x_i)\right):i=1,\ldots,n\right\}.$$
We first assume that the underlying copula is Archimedean and
independent of \ttf{CHOL}: it is specified using the new spline
expression proposed in (\ref{lambert:NonparametricGenerator}) with
$K=11$ B-splines in the basis and a 3rd order
penalty. Figure~\ref{FraminghamFig1} illustrates the smoothness of the
fitted joint distribution (see the improvement over Fig.~5 in
\citet{Lambert:2007vo}) and suggests that the Gumbel copula is a good
parametric approximation to the dependence structure underlying the
log-blood pressures.
Conditional copulas were also fitted using the methodology described
in Sections \ref{AdditiveCondSplineFamily:Sect} and
\ref{FlexPowerFamily:Sect} with the same parameters as in the first
application. A comparison of the DIC values for the unconditional
copula model and for the conditional ones (see Table
\ref{BloodPressures:DIC:Tab}) suggests that the strength of
association between blood pressures is not changing with the
cholesterol level.
\begin{table}
\centering
\begin{tabular}{lccc}
Copula model & \# par. & e.d. & DIC \\
\hline
Model 0 (Unconditional) & 11 & 4.2 & -569.0 \\
Model 1 (additive)      & 16 & 5.2 & -567.9 \\
Model 2 (flex-power)    & 21 & 5.1 & -567.5
\end{tabular}
\wcaption{Blood pressure dataset: effective number of parameters (e.d.) and DIC in the fitted copula models.}{BloodPressures:DIC:Tab}
\end{table}
This can be visualized on Fig.~\ref{BloodPressures:FittedTau:Fig}
where the simultaneous credible regions for the conditional Kendall's
tau in the additive and in the flex-power Archimedean copula families
are superposed to the HPD interval $(0.52,0.58)$ for $\tau$ in the
unconditional copula model.

\section{Discussion}
A performant spline estimator for the generator of an Archimedean
copula has been proposed. A large simulation study has shown that it
has nearly negligible bias and credible regions with coverage
probabilities closed to their nominal values even for moderate sample
sizes.

It has been extended to change smoothly with a covariate using either
an additive model for the spline coefficients or flexible (spline
based) power transforms of the generator and of its argument. Again,
simulation studies suggest that these models are quite flexible and
useful to describe changes in conditional association structures. 

It can be extended to handle multiple categorical or continuous
covariates using additive forms such as in
(\ref{derivativeOfCondG4:Eq}). More than two responses can also be
considered using conditional copulas of the form
\begin{eqnarray*}
C(u_1,\ldots,u_J|x) = \varphi^{-1}\left({
\varphi(u_1|x)+\ldots+\varphi(u_J|x)}|x\right),
\end{eqnarray*}
but one should question the validity of an exchangeable dependence
structure in the specific modelling exercise. 

A joint estimation of the parameters involved in the marginal models
and in the conditional copula structure is also possible and
straightforward to manage in a Bayesian framework. Unless at least one
of the margins is discrete, it is not obvious that one would gain
anything substantial by merging these modelling efforts in a single
inferential procedure.

Besides additive models, an extension to hierarchical settings
involving repeated measurements (such as with clustered or
longitudinal data) is certainly worth considering, but out of scope
for this paper. The proposed conditional copula models can also be
used as building block in pair copula constructions such as in regular
vines \citep[see e.g.][]{Aas:2009fga}.


\section*{Acknowledgments}
The author acknowledges financial support from IAP research network
P7/06 of the Belgian Government (Belgian Science Policy), and from the
contract `Projet d'Actions de Recherche Concert\'ees' (ARC) 11/16-039
of the `Communaut\'e fran\c{c}aise de Belgique', granted by the
`Acad\'emie universitaire Louvain'.

\appendix
\section{Likelihood  in the spline copula model} \label{Appendix:1}
From a computational point of view, we found convenient to rewrite the
likelihood as
\begin{align}
 & L({\vec\theta}) = -\prod_{i=1}^n \left( 1 -\lambda'_{\vec\theta}(C_i)\right)
 {\lambda_{\vec\theta}(C_i) \over \lambda_{\vec\theta}(u_i) \lambda_{\vec\theta}(v_i)} \,
 {\varphi_{\vec\theta}(u_i)\, \varphi_{\vec\theta}(v_i) \over 
  \left({\varphi_{\vec\theta}(u_i) +
      \varphi_{\vec\theta}(v_i)}\right)^2}
\label{Likelihood:computation:Eq}
\end{align}
where
$$\lambda_{\vec\theta}(u) = {\varphi_{\vec\theta}(u) \over \varphi'_{\vec\theta}(u)}
= {-1 \over g_{\vec\theta}'(S(u)) S'(u)}.$$
The value of $C_i$ is given by
\begin{align*}
& C_i = \varphi_{\vec\theta}^{-1}(\varphi_{\vec\theta}(u_i)+\varphi_{\vec\theta}(v_i)).
\end{align*}
The function $\varphi_{\vec{\theta}}^{-1}(\cdot)$ can be computed
numerically using an iterative method. The following algorithm turns
to be very efficient. Assume that one looks for $u$ such that
$\varphi^{-1}_{\vec\theta}(x)=u$. This is equivalent to finding the
root of $f(\cdot)$ with $f(u) = g_{\vec{\theta}}(S(u)) +
\log(x)$. Starting from $u=u^{(0)}$, the guess $u^{(t+1)}$ at iteration
$t+1$ (resulting from a Newton step) is such that $S(u^{(t+1)}) =
S(u^{(t)}) + \delta^{(t)}$ with $\delta^{(t)}= -f(u^{(t)}) /
g'(S(u^{(t)}))$. Starting from $C_{i}^{(0)}=u_{i}v_{i}$ and letting
$x_i=\varphi_{\vec\theta}(u_i)+\varphi_{\vec\theta}(v_i)$, that
procedure can be vectorized to compute $C_i$ ($i=1,\ldots,n$) with a
single loop in three or four iterations.  Finally, note that the
quantities $\lambda'_{\vec\theta}(C_i)$ in the likelihood can be
approximated using finite differences of the lambda function.



\end{document}